%% file: main.tex
\documentclass[acmsmall,screen]{acmart}
\AtBeginDocument{%
  }

\setcopyright{acmlicensed}
\copyrightyear{2025}
\acmYear{2025}
\acmDOI{XXXXXXX.XXXXXXX}

\acmJournal{JACM}
\acmVolume{37}
\acmNumber{4}
\acmArticle{111}
\acmMonth{8}

\input{macros}

\begin{document}

\title{\SystemName: Automated Dataset Description Generation using Large Language Models}

\author{Haoxiang Zhang}
\authornote{Equal contribution.}
\email{haoxiang.zhang@nyu.edu}
\orcid{1234-5678-9012}
\affiliation{%
  \institution{New York University}
  \city{New York}
  \state{NY}
  \country{USA}
}

\author{Yurong Liu}
\authornotemark[1]
\email{yurong.liu@nyu.edu}
\orcid{1234-5678-9012}
\affiliation{%
  \institution{New York University}
  \city{New York}
  \state{NY}
  \country{USA}
}

\author{A\'{e}cio Santos}
\affiliation{%
  \institution{New York University}
  \city{New York}
  \state{NY}
  \country{USA}
}
\email{aecio.santos@nyu.edu}

\author{Wei-Lun (Allen) Hung}
\affiliation{%
  \institution{New York University}
  \city{New York}
  \state{NY}
  \country{USA}
}
\email{allen.hung@nyu.edu}

\author{Juliana Freire}
\affiliation{%
  \institution{New York University}
  \city{New York}
  \state{NY}
  \country{USA}
}
\email{juliana.freire@nyu.edu}

\renewcommand{\shortauthors}{Zhang, Liu, et al.}

\begin{abstract}
\input{sections/Sec0_Abstract}
\end{abstract}




\maketitle

\input{sections/Sec1_Introduction}

\input{sections/Sec3_ProblemDefinition}

\input{sections/Sec3_System}

\input{sections/Sec4_Benchmark}

\input{sections/Sec5_Experiment}

\input{sections/Sec2_RelatedWork}

\input{sections/Sec6_Conclusion}

\bibliographystyle{ACM-Reference-Format}
\bibliography{paper}

\newpage
\arxiv{
\appendix
\input{sections/Sec7_Appendix}
}

\end{document}

%% file: macros.tex
\usepackage{xspace}
\usepackage{xcolor}

\usepackage[most]{tcolorbox}

\usepackage{algorithm}
\usepackage{algcompatible}

\usepackage{graphicx}
\usepackage{subcaption} 
\usepackage{enumitem}

\usepackage{xcolor}
\usepackage{soul}

\newcommand{\myparagraph}[1]{\noindent{\textbf{#1.}}~}

\newcommand{\revshep}[1]{#1}

\def\withrevision{0} 
\ifnum\withrevision=1

    \newcommand{\revone}[1]{\textcolor{blue}{#1}}
    \newcommand{\revtwo}[1]{\textcolor{cyan}{#1}}
    \newcommand{\revthree}[1]{\textcolor{teal}{#1}}
    \newcommand{\revgen}[1]{\textcolor{violet}{#1}}
\else

    \newcommand{\revone}[1]{#1}
    \newcommand{\revtwo}[1]{#1}
    \newcommand{\revthree}[1]{#1}
    \newcommand{\revgen}[1]{#1}
\fi

\newcommand{\hide}[1]{}

\def\withnotes{1} 
\ifnum\withnotes=1
    \newcommand{\ah}[1]{\textcolor{violet}{Allen: #1}}
    \newcommand{\yl}[1]{\textcolor{brown}{Yurong: #1}}
    \newcommand{\as}[1]{\textcolor{cyan}{Aecio: #1}}
    \newcommand{\jf}[1]{\textcolor{red}{Juliana: #1}}
    \newcommand{\hz}[1]{\textcolor{blue}{Haoxiang: #1}}
\else
    \newcommand{\ah}[1]{\xspace}
    \newcommand{\yl}[1]{\xspace}
    \newcommand{\as}[1]{\xspace}
    \newcommand{\jf}[1]{\xspace}
    \newcommand{\hz}[1]{\xspace}
    \renewcommand{\revone}[1]{#1}
    \renewcommand{\revtwo}[1]{#1}
    \renewcommand{\revthree}[1]{#1}
    \renewcommand{\revgen}[1]{#1}
\fi

\def\withnotes{1} 
\ifnum\withnotes=0
    \newcommand{\fr}[1]{\textcolor{red}{Future/Revision: #1}}
    
\else
  \newcommand{\fr}[1]{\xspace}
  
\fi

\newcommand{\SystemName}{AutoDDG\xspace}

\usepackage{cleveref}
\definecolor{HighlightColor}{rgb}{0.05,0.05,0.70}
\def\submission{0}

\ifnum\submission=1
    \newcommand{\subm}[1]{#1}
    \newcommand{\arxiv}[1]{\xspace}
\else
    \newcommand{\arxiv}[1]{#1}
    \newcommand{\subm}[1]{\xspace}
\fi

\newcommand{\savespace}[1]{}
\newcommand{\revised}[1]{\textcolor{HighlightColor}{#1}\xspace}
\renewcommand{\revised}[1]{{#1\xspace}}

\newcommand{\PromptFontSize}{%
  \fontsize{7.5}{9}\selectfont
}

%% file: sections/Sec0_Abstract.tex
The proliferation of datasets across open data portals and enterprise data lakes presents an opportunity for deriving data-driven insights. Widely-used dataset search systems rely on keyword search over dataset metadata, including descriptions, to support discovery. Therefore, when these descriptions are incomplete, missing, or inconsistent with dataset contents, findability is severely compromised.
To improve findability,  we introduce \SystemName, a framework that automatically generates descriptions of tabular data. 
By adopting a data-driven approach to summarize dataset contents and leveraging large language models (LLMs) to enrich summaries with semantic information and produce human-readable text, \SystemName derives descriptions that are comprehensive, accurate, readable, and concise.
A critical challenge in this problem is evaluating the effectiveness of description generation methods and assessing the quality of the generated descriptions. We propose a comprehensive evaluation methodology that combines retrieval, reference-based, and reference-free assessment, with human validation. 
Our experimental results using new benchmarks demonstrate that \SystemName generates high-quality, accurate descriptions at scale, significantly improving dataset retrieval performance across diverse use cases.
\SystemName is publicly available at https://github.com/VIDA-NYU/AutoDDG.

%% file: sections/Sec1_Introduction.tex
\vspace{-0.15cm}
\section{Introduction}
\label{sec:intro}

We have witnessed a proliferation of data portals and data lakes~\cite{gregory2022human, figshare, zenodo, dataverse, hendler2012datagov, kassen2013opendata_chicago, NYC_opendata,nargesian2019data} as more \revone{structured} data is generated and made available. It is estimated that there are tens of millions of datasets on the web~\cite{google-dataset-analysis}.
\textbf{However, a significant fraction of these datasets remain effectively invisible due to inadequate or missing descriptions~\cite{koesten2017trials}.}
While these datasets present significant opportunities for data-driven insights, they must be easily discoverable and interpretable to maximize their utility~\cite{wilkinson2016fair}.

\begin{figure}[t!]
  \includegraphics[width=\textwidth]{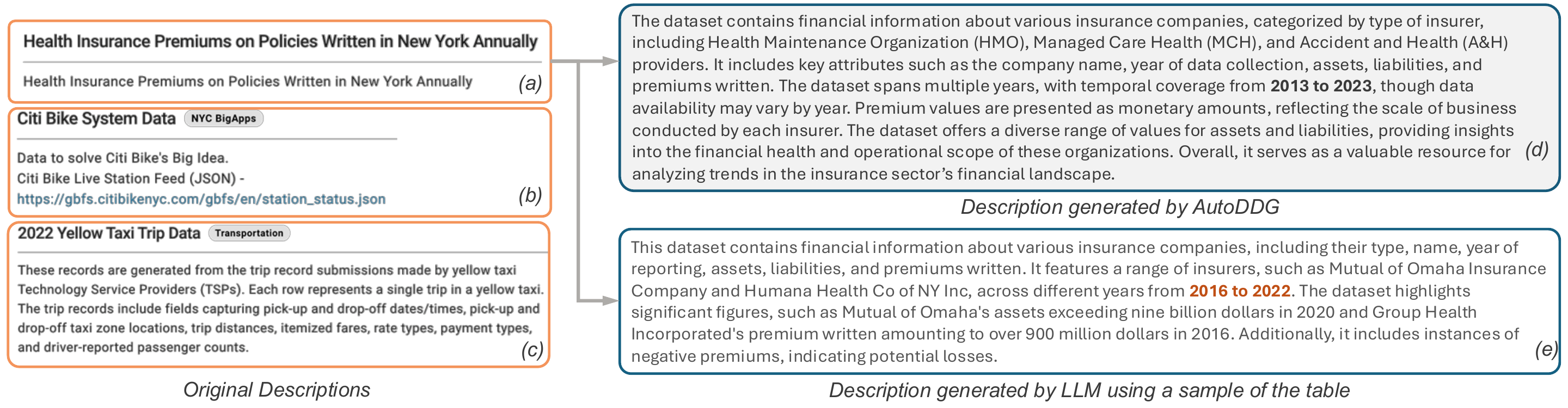}
  \caption[]{\revthree{Dataset descriptions from NYC Open Data: (a)  and (b) lack sufficient details, while (c) contains inconsistent information--the \textit{2022 Yellow Taxi Trip Data} dataset includes taxi trips that span multiple years, not just 2022. (d) shows a description automatically generated by \SystemName for the dataset \texttt{(a)}, while (e) was generated by an LLM using a data sample.}}
  \label{fig:example_original_generate_description}
\end{figure}

Dataset search systems that power data portals~\cite{brickley2019google, CKAN, socrata} follow the web search paradigm: they support keyword queries over metadata, such as dataset names and descriptions. Consequently, discoverability depends on metadata quality--how effectively it conveys the dataset contents and aligns with users’ information needs.
The critical importance of descriptions is exemplified by Google Dataset Search, which excludes from its index datasets lacking descriptions~\cite{benjelloun2020google}. Platforms such as CKAN do not mandate descriptions but encourage their use to improve search precision and recall, facilitate dataset understanding, and help users determine relevance~\cite{CKAN_description}.
Yet many datasets are published with inadequate or missing descriptions. In the index of the Auctus dataset search engine \cite{castelo2021auctus},  3,121 out of 23,520 datasets (13.2\%) have no descriptions, and 2,346 datasets (approximately 10\%) have descriptions with 10 words or fewer. 
\autoref{fig:example_original_generate_description} (a) and (b) show dataset descriptions from NYC Open Data~\cite{NYC_opendata} that are terse and fail to convey key characteristics of the dataset; 
%
(c) is actually \textit{inconsistent} with the actual contents--it says that the dataset includes taxi trips from 2022, when the actual data contains records that span multiple years.

Datasets without detailed and accurate descriptions may be technically available, but they are effectively invisible. \citet{koesten2017trials} compare the problem of finding data today to the early days of the web, when users needed to know the URL of web pages or relied on manually created directories such as DMOZ to access content.
Beyond discoverability, descriptions are crucial for relevance assessment. Just as web search users rely on snippets to decide which results to explore, dataset search users depend on descriptions to filter relevant datasets. This issue is particularly pronounced for datasets, where downloading and inspecting large files is far more costly than simply opening a webpage.

\footnotetext[1]{The \textit{Health Insurance} dataset:\url{https://data.ny.gov/Economic-Development/Health-Insurance-Premiums-on-Policies-Written-in-N/xek8-zfrt/about_data}}
\footnotetext[2]{The \textit{Citi Bike} dataset: \url{https://data.cityofnewyork.us/NYC-BigApps/Citi-Bike-System-Data/vsnr-94wk/about_data}}
\footnotetext[3]{The \textit{Yellow Taxi} dataset: \url{https://data.cityofnewyork.us/Transportation/2022-Yellow-Taxi-Trip-Data/qp3b-zxtp/about_data}}

\myparagraph{LLMs for Automatic Description Generation}
The ability of language models to produce fluent, readable text, combined with their broad world knowledge~\cite{brown2020gpt,touvron2023llama}, enables them to generate coherent descriptions enriched with inferred semantic and contextual information. Prior work has demonstrated the effectiveness of LLMs in semantic inference tasks for tabular data, including the ability to determine the semantic types of attributes and the table class~\cite{archetype@vldb2024,chorus@vldb2024}. By leveraging external knowledge not explicitly encoded in the data, LLMs can enhance dataset descriptions in ways that manual approaches might overlook.

However, using LLMs for this task presents several challenges. 
LLMs are designed for processing and generating textual data, whereas \revone{datasets are structured in tabular form}. This mismatch raises questions about how to best represent tables in prompts, a challenge that remains an active area of research in table representation learning~\cite{sui2024table}. Moreover, fixed context windows limit LLMs' ability to process large datasets in their entirety~\revone{\cite{li2024longcontextllmsstrugglelong,chorus@vldb2024,archetype@vldb2024}}. Relying on data samples risks yielding incomplete or misleading descriptions that fail to capture crucial global properties such as the spatial and temporal coverage of the dataset. \revone{This is illustrated in Figure~\ref{fig:example_original_generate_description}(e): the description derived by an LLM using a sample states that the data covers the years 2016-2022, while it actually includes records from 2013-2023.}

\revone{
Furthermore, LLMs can generate hallucinated or inaccurate content \cite{huang2025survey, watanabe2024capabilities}, making it critical to implement safeguards to ensure that descriptions remain grounded in the dataset's contents. 
}

\begin{figure*}
  \centering
  \includegraphics[width=\textwidth]
  {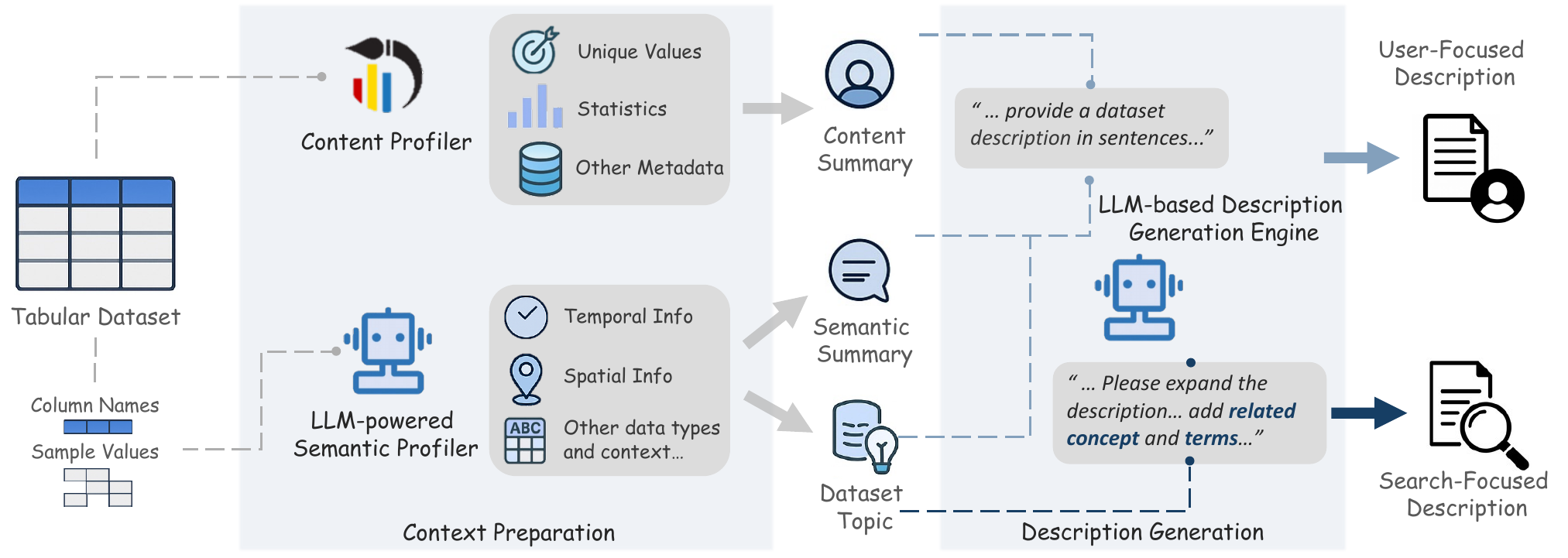}
  \vspace{-0.3cm}
  \caption{
  \textit{\SystemName: a multi-stage framework for tabular dataset description generation.}
    In the \textit{Context Preparation} stage, a content and a semantic profiler derive summaries of the dataset.
    %
    The LLM-powered \textit{Description Generation Engine} uses the summaries to produce dataset descriptions. The system can generate descriptions tailored to specific needs.
    %
    %
  }
  \label{fig:solution_overview}
  \vspace{-0.3cm}
\end{figure*}

\revshep{ \noindent 
\myparagraph{Limitations of Existing Approaches}
The main limitation of existing methods is their inability to synthesize comprehensive and accurate descriptions of heterogeneous datasets that are common in data lakes and open data repositories. 
Pre-trained table-to-text models, such as MVP~\cite{tang2023mvp} and ReasTAP~\cite{zhao2022reastap}, were designed for domain-specific narrative tasks, including sports summaries and biographical sentences. When applied to diverse datasets, these models exhibit poor generalization and generate descriptions that do not align with users’ needs for discoverability. 
Simpler heuristics, such as concatenating column headers with sample values (Header+Sample), and LLM-only approaches, rely solely on data samples for context generation. This reliance on samples risks producing incomplete or misleading descriptions that fail to capture crucial global dataset properties, such as the full temporal extent or statistical ranges across all rows. Furthermore, as illustrated above, the absence of grounding in global statistics can lead LLM-only methods to hallucinate, resulting in descriptions with factual errors.
Pneuma~\cite{pneuma@sigmod2025} leverages LLMs for schema narration and row samples to generate table summaries for data discovery.
Since it uses row samples, it misses the global properties of columns. Additionally, Pneuma produces dense representations designed primarily for retrieval rather than human readability, limiting their utility for users assessing dataset relevance.
}

\myparagraph{Our Approach: \SystemName}
\revshep{To address these limitations,} we introduce \SystemName, a \revone{systematic LLM-powered framework for the automatic generation of tabular dataset descriptions}.
\revone{Instead of directly feeding data samples to LLMs, which risks missing global properties and generating hallucinations, \SystemName employs a two-stage architecture (Figure~\ref{fig:solution_overview}):}

\noindent \revone{\textit{Context Preparation:} We profile datasets using two complementary approaches 
(1) \textit{Content profiling} uses algorithmic techniques to capture global statistical properties~\cite{castelo2021auctus,naumann2014data}, ensuring descriptions are grounded in actual data characteristics and fit within LLM context windows;
(2) \textit{Semantic profiling} leverages LLMs to enrich summaries with contextual information not explicitly present in the data, such as the dataset topic and types of analyses the dataset can be used for, improving discoverability through meaningful keywords and topical cues.} 

\noindent  \revshep{\textit{Description Generation:} An LLM-based engine uses these summaries to generate two types of descriptions that balance conciseness and comprehensiveness: User-Focused Descriptions (UFD) optimized for readability, providing succinct yet informative overviews; and Search-Focused Descriptions (SFD), which incorporate dataset overviews, themes, and keyword-rich snippets for improved retrieval (Section~\ref{sec:solution_UFD_SFD}).}
\revone{These descriptions can be indexed by standard search engines (e.g., Lucene~\cite{lucene}, Elasticsearch~\cite{elasticsearch}, and Solr~\cite{apachesolr}) \revshep{as well as used to create embeddings} to support the efficient evaluation of textual queries, thus improving discoverability without requiring major infrastructure changes.}
\revshep{Unlike existing approaches, AutoDDG does not require training and generalizes to different domains. It derives descriptions that are grounded in dataset statistics,
enriched with contextual information, and human-readable. As we demonstrate in Section~\ref{sec:experiment}, the AutoDDG descriptions  are 1) highly effective for retrieval;
2) accurate--as judged by expert evaluators; and
3) readable--attaining high readability scores from both LLM judges and expert evaluators. }

By using summaries, \SystemName produces descriptions with consistent structure, enhancing the usability of dataset search tools and
addressing a common user frustration with encountering metadata that varies unpredictably across search results~\cite{Sostek2024Discovering}. 
\revshep{Additionally, AutoDDG can complement end-to-end data discovery systems such as Pneuma~\cite{pneuma@sigmod2025} by providing high-quality dataset descriptions that improve both retrieval effectiveness and human readability.}

%
%




\revgen{
We introduce a comprehensive
evaluation methodology to assess the quality of generated descriptions that measures: 
1)  the improvement in dataset retrieval when automatically generated descriptions are used,
2) the similarity between the generated descriptions and existing ones (when available) to evaluate the ability of AutoDDG to produce descriptions that resemble human-generated descriptions, 
(3) intrinsic text quality metrics, using LLMs as a judge~\cite{liu2023geval, gao2024llmEval, seo2024unveiling, zhao2023investigating} (\revthree{and validated by human experts}) to evaluate language-level aspects such as readability and conciseness (Section~\ref{sec:solution_evaluation}),
(4) the cost and scalability of the approach.
}
%
%
%
%
%

\myparagraph{Summary of Contributions}
\revone{This paper presents the first systematic LLM-powered framework for automated dataset description generation to improve dataset discoverability and relevance assessment in keyword-based search engines.}
%
%
%
%
Our main contributions can be summarized as follows:
\begin{itemize}[leftmargin=*,noitemsep]
\vspace{-.1cm}
\item \revgen{\noindent\textit{The \SystemName Framework:} A dual-profiling architecture that addresses fundamental systems challenges—processing large datasets within LLM context limits while maintaining data faithfulness and enriching the descriptions with semantic information.  
\revshep{\SystemName generates descriptions optimized for different goals--readability and retrieval.}
}
%


\item \revgen{\textit{Benchmarks and Evaluation Methodology:} We propose a multi-pronged approach to evaluate different aspects of descriptions, combining reference-based, reference-free, and expert assessment. We also construct two benchmarks, ECIR-DDG and NTCIR-DDG, tailored to evaluate the quality of dataset descriptions for keyword-based search and retrieval tasks. }

\item \revgen{\textit{Experimental Evaluation:} We report the results of an extensive experimental evaluation that validates our design decisions and demonstrates AutoDDG's effectiveness at generating descriptions that improve dataset retrieval up to 30\% compared to the original descriptions, 
with robust performance across commercial and open-weight LLMs.
\revshep{
We also propose two optimized versions of \SystemName that improve efficiency and assess its cost and scalability relative to other dataset description generation methods.}
}

\end{itemize}

%% file: sections/Sec3_ProblemDefinition.tex
\vspace{-.15cm}
\section{\SystemName: LLM-Powered Automated Dataset Description Generation}
\label{sec:auto_ddg}

\myparagraph{Problem Definition}
Given a tabular dataset $D$ with columns $C$ and rows $R$, our task is to \textit{automatically} generate a descriptive \revone{textual} summary $S_D$ that effectively captures the key characteristics of~$D$, such as column names, data types, statistical properties, and semantically enriched information, \revised{to support dataset discoverability and relevance assessment.}
\revone{
Formally, our goal to design a function $\mathcal{G}(\cdot)$ that takes as input the dataset $D$ and generates a textual description $S_D$:
\vspace{-.15cm}
\[ S_D = \mathcal{G}(D) 
\]
In this paper, we focus on the subclass of functions $\mathcal{G}$ that relies on LLMs to generate descriptions. Implementing such a function is non-trivial due to the large design space for both the input context provided to the LLM and the instructions necessary to shape the desired output.
In practice, this entails designing a set of prompt templates $P$, a dataset context $X(D)$, as well as algorithms to orchestrate the generation process.
More specifically, our problem involves designing effective methods for:
}

\noindent \textit{Description Generation}: Crafting prompts $P$ that guide the LLM to generate accurate and rich descriptions tailored to specific use cases, such as enhancing human readability or discoverability.

\noindent\textit{Context Preparation}: Extracting information and representing the dataset context $X(D)$ in a way that captures the heterogeneous characteristics of tabular data, fits within the input limitations of LLMs, and is aligned with users' information needs.
\revtwo{
In this paper, we limit \SystemName to use only the table content and the information automatically derived from it (e.g., using LLMs and data profilers) to create the context $X(D)$.
We deliberately exclude existing metadata to assess the framework's ability to generate descriptions from data alone, addressing the common practice of datasets published with inadequate or no metadata. However, when additional metadata is available (e.g., data provenance or existing categorizations), it can be incorporated as context. This is a direction that we intend to pursue in future work. }

%
%
%


%

%
\myparagraph{Desiderata for Dataset Descriptions}
\revised{
We ground our design decisions on several studies on information-seeking behavior and data discovery.
These studies have revealed a gap between the datasets available and those that users can effectively locate~\cite{chapman2020dataset}. 
They have also highlighted shortcomings in existing dataset metadata that hinder discoverability and the assessment of relevance~\cite {papenmeier2021genuine, koesten2017trials, Sostek2024Discovering}. Building on these findings, which we summarize in \Cref{sec:related_work}, we propose a set of desiderata for dataset descriptions:
}
\begin{itemize}[itemindent=0.5cm,leftmargin=0.0cm,topsep=0.25em]
    \item \emph{Faithfulness to Data:} Dataset descriptions must be accurate representations of the dataset contents.
    \item \emph{Uniformity:} While variability is expected in dataset collections, it is important to maintain a certain uniformity in the descriptions to facilitate comparison and relevance assessment.
    \item \emph{Conciseness and Readability:} Dataset descriptions should be concise to enable quick relevance assessment and yet detailed enough to convey meaningful insights. 
    \item \emph{Comprehensiveness and Alignment with Users' Information Needs:} Widely-used search systems~\cite{CKAN,socrata,brickley2019google} often do not account for dataset contents, limiting users' ability to locate relevant data. Thus, descriptions should provide details about the contents, ensuring that they address users' needs.
\end{itemize}

\myparagraph{Discussion} \revised{
There is an inherent trade-off between conciseness and comprehensiveness: overly brief descriptions may fail to capture key aspects of the dataset, whereas excessively detailed descriptions can be difficult and time-consuming for users to process. Furthermore, while readability enhances users’ ability to assess relevance, it is less critical for indexing and automated discoverability.
}

\revised{
To achieve a balance between these conflicting goals, we design our framework to generate multiple outputs tailored to different objectives. In particular, our framework currently generates descriptions to support (1)  users' understanding of dataset contents and (2) building indexes for dataset search engines. These descriptions prioritize, respectively, readability and comprehensiveness.}

%% file: sections/Sec3_System.tex
\begin{figure}
    \centering
    \textsf{\small Content summary for ``Health Insurance Dataset''.}
    \vspace{-.75em}
    \begin{tcolorbox}[colback=black!2.5!white,colframe=black!85!black,boxrule=0.25mm,boxsep=4pt,left=0pt,right=0pt,top=0pt,bottom=0pt]
    \scriptsize
    \tt
    \begin{verbatim}
Number of Rows: 790
Number of Columns: 6
Columns:
  - Name: Year
    - Data Types: Text, DateTime
    - Coverage: 2013 to 2022
    - Unique Values: 10
  - Name: Liabilities
    - Data Types: Integer
    - Coverage: 0 to 2682301090.0
    - Unique Values: 757
  ...
  [other attributes are omitted due to space limitations]
    \end{verbatim}
    \vspace{-1.5em}
    \end{tcolorbox}

    \vspace{-1.5em}
    \caption{Example summary created by the Content Profiler.}
    \label{fig:example_content_profile}
\end{figure}


\subsection{Context Preparation}
\label{sec:solution_STA}


\revised{The context preparation step takes as input a table and profiles it, generating \textit{dataset summaries} as outputs. We currently implement two profilers: the \textit{Content Profiler} (Section \ref{sec:content-profiler}) and the Semantic Profiler (Section \ref{sec:semantic-profiler}). The output of these profilers is combined to create the \textit{dataset context}, which is used to generate the description.}

\subsubsection{Content Profiler} 
\label{sec:content-profiler}
\revised{The summary generated by the \textit{Content Profiler} is derived directly from the table's contents.
It captures structural and statistical properties that are commonly extracted by traditional data profilers~\cite{abedjan2015profiling, naumann2014data,datamart-profiler}.
In our implementation, we use the Datamart Profiler~\cite{datamart-profiler}. 
%
Each table is profiled by analyzing all rows of each column to extract information such as attribute data types, value distributions, and uniqueness.
\autoref{fig:example_content_profile} shows an example of such a content summary.
These elements provide a global view of a dataset, summarizing its structure and content in a concise manner that can be effectively fed into LLMs.
Beyond statistical summaries, content profiling could be extended to domain-specific data (e.g., from the biomedical domain) to extract metadata that aligns with specific application needs.
}

\subsubsection{Semantic Profiler}
 \label{sec:semantic-profiler}
\revised{ 
The \textit{semantic summary} goes beyond the dataset contents to capture contextual information.
\SystemName uses LLMs to enrich the dataset context with external knowledge that complements the \revised{content summary} and caters to users' information needs.
This includes the dataset topic and semantic information about all columns in the table and their potential uses, which improves interpretability and relevance~\cite{koesten2020everything}.}

\begin{algorithm}[t]
  \caption{SemanticProfiler}
  \label{alg:semantic_type_analyzer}
\begin{algorithmic}[1]
  \STATE {\bfseries Input:} Tabular Dataset $D$, LLM model $M$, Number of Samples $sample\_size$
  \STATE {\bfseries Output:} Semantically enriched information for each column in $D$
  \STATE Initialize list $semantic\_summary$ as empty
  \FOR{\textbf{each} column $C_i$ in $D$}
      \STATE $sample\_values$ = GetSample($C_i$, $sample\_size$)
      \STATE $semantic\_info$ = Prompt($M$, $C_i$, $sample\_values$)
      \STATE Create a human-readable summary $column\_summary$ based on the semantic information $semantic\_info$
      \STATE Append $column\_summary$ to $semantic\_summary$
  \ENDFOR
  \STATE \textbf{return} $semantic\_summary$
\end{algorithmic}
\end{algorithm}








Algorithm~\ref{alg:semantic_type_analyzer} describes the structured prompting approach we use to guide the LLM.
For each column, it generates a prompt that includes the column name, sample values, and data type. The LLM responds with a JSON-formatted classification of the column based on predefined semantic categories.
\revised{Examples of the output of the structure-defined prompt are shown in \autoref{fig:examples_semantic_profiler}. 
%
The structure-defined template and the complete prompt for semantic enrichment analysis are available in 
%
\subm{the extended version of this paper \cite{AutoDDG_arxiv}.}
\arxiv{Tables~\ref{tab:template} and \ref{tab:prompt_semantic_profile}  in the Appendix.} 
\revshep{Note that AutoDDG supports different execution modes, including a high-concurrency mode for speed and a group-prompting mode for token efficiency; we discuss these in Section~\ref{sec:experiment}.}
}

\input{tables/example_sp}

The structured output is then serialized into descriptive sentences, and the summaries of all columns are concatenated to form the final output of the semantic profiler module.
For example, the serialized semantic summary for Example 1 (from \autoref{fig:examples_semantic_profiler}) is:\\
\texttt{\small **Year**: Represents temporal entity. Contains temporal data (resolution: Year).} \\
\texttt{\small Domain-specific type: general. Function/ Usage Context: Aggregation Key.}\\
These enriched outputs provide a detailed understanding of each column's semantic properties, enabling the generation of high-quality, contextually relevant dataset descriptions. 

To enhance dataset discoverability, we also incorporate dataset topics into the semantic summary. 
The process leverages LLMs to analyze dataset metadata and samples to extract meaningful topics that provide a high-level overview by capturing the primary theme of a dataset in 2-3 words.
This entails two key steps: (1) \textit{prompt design}, where a tailored prompt is dynamically constructed using the dataset's title, original description (if available), and sample data; (2) \textit{topic generation}, where the LLM processes the prompt and generates a brief topic. \autoref{tab:dataset_topic_prompt} shows the prompt used by the dataset topic generator.

\myparagraph{Discussion}
\revised{
The design of our prompts was guided by the findings of studies on information seeking requirements for dataset search (summarized in Section~\ref{sec:related_work}).
}
Our goal is to obtain information that is aligned with how users formulate search queries. For example, the semantic profiler includes information about temporal and spatial attributes, as well as their resolution, as this was a common pattern observed by~\citet{koesten2017trials}.
The semantic profiler (and corresponding prompts) can be adapted to support other information needs and domains. It can also be extended to include other useful information, such as other semantic types of interest~\cite{archetype@vldb2024,chorus@vldb2024}.

\begin{figure}
    \small
    \PromptFontSize
    \centering
    \begin{tabular}{p{\textwidth}}
    \toprule
    \textbf{\small Dataset Topic Generation Prompt} \\
    \midrule
    \texttt{Using the dataset information provided, generate a concise topic in 2-3 words that best describes the dataset's primary theme:} \\
    \texttt{- Title: \textcolor{purple}{\{title\}} } \\
    \texttt{- Original Description: \textcolor{purple}{\{original\_description\}} (optional)} \\
    \texttt{- Dataset Sample: \textcolor{purple}{\{dataset\_sample\}}} \\
    \texttt{- Topic (2-3 words):} \\
    \bottomrule
    \end{tabular}
    
    \vspace{-1em}
    \caption{Prompt for generating concise dataset topics based on the dataset title, description, and sample data.}
    \label{tab:dataset_topic_prompt}
\end{figure}

\subsection{Description Generation}
\label{sec:solution_UFD_SFD}

\SystemName uses LLMs to generate two types of dataset descriptions: User-Focused Descriptions (UFD) and Search-Focused Descriptions (SFD). For this task, we carefully design prompts to guide the LLM toward producing descriptions optimized for dataset search engine users to assess relevance and to improve dataset discoverability.

\subsubsection{User-Focused Description (UFD)}
The UFD is designed to provide a clear, concise, and accurate overview of the dataset, prioritizing human readability and presentation. This type of description works best for scenarios where the dataset needs to be communicated to users in a way that is easily understood, such as in reports, dashboards, or data catalogs aimed at human readers. Although its primary goal is readability, UFD can also be effective for search purposes, as it offers a well-structured overview containing key terms and concepts related to the dataset. The UFD is generated by prompting the LLM to describe the dataset based on dataset samples and the content summary (Section~\ref{sec:solution_STA}).
The UFD prompt and a sample output are shown in \autoref{tab:prompt_example_ufd}.
%
The description includes information about the temporal extent and resolution, number of records, as well as
a summary of the contents (e.g., the range of wind speed values)
and an overview statement about the dataset and what it can be used for. 

\begin{figure}
    \small
    \centering
    \begin{tabular}{p{\textwidth}}
    \toprule
    \textbf{User-Focused Description Prompt} \\
    \midrule
    \PromptFontSize
    \texttt{Answer the question using the following information. First, consider the dataset sample: \textcolor{purple}{\{D\_sample\}}. Additionally, the dataset profile is as follows: \textcolor{purple}{\{D\_profile\}}. Based on this profile, please add sentence(s) to enrich the dataset description. Furthermore, the semantic profile of the dataset columns is as follows: \textcolor{purple}{\{D\_semantic\}}. Based on this information, please add sentence(s) discussing the semantic profile in the description. Moreover, the dataset topic is: \textcolor{purple}{\{D\_topic\}}. Based on this topic, please add sentence(s) describing what this dataset can be used for.}
    \texttt{Based on the information above and the requirements, provide a dataset description in sentences. Use only natural, readable sentences without special formatting.}\\
    \bottomrule
    \end{tabular}

    \vspace{1em}
    \textsf{Example Output:}
    \vspace{-.5em}
    \begin{tcolorbox}[colback=black!2.5!white,colframe=black!85!black,boxrule=0.25mm,boxsep=4pt,left=0pt,right=0pt,top=0pt,bottom=0pt]
    \PromptFontSize
    \tt
    This dataset contains wind speed and direction measurements from a specific time period in 2003. The data includes 4433 unique time stamps, with a temporal coverage of May 13 to June 12, 2003, at a resolution of minutes. The average wind speed ranges from 0 to 27.0, with a standard deviation ranging from 0 to 2.78. The average wind direction ranges from 0 to 338.0, with 16 unique values. The dataset provides a comprehensive view of wind patterns during this time period, making it suitable for environmental studies and research.
    
    \end{tcolorbox}

    \vspace{-.5cm}
    \caption{Prompt and example of a UFD.}
    \label{tab:prompt_example_ufd}
\end{figure}

\subsubsection{Search-Focused Description (SFD)}

The SFD is optimized to enhance the discoverability of datasets in search systems. \revised{The process begins with a tabular dataset, which is processed by an LLM to generate an initial description and identify the dataset topic.}
By including a specific topic or area related to the dataset, the LLM can focus on expanding the description with relevant terms, concepts, synonyms, and keyword variations. This helps improve search engine indexing and retrieval performance.
For example, suppose a user wishes to publish a dataset on "climate data" that contains detailed measurements across various regions. The system 
identifies "climate data" as the topic and 
%
%
enhances the description by including climate-related keywords such as "temperature trends," "precipitation," "regional climate analysis," and "weather patterns." This process enables the SFD to incorporate terms related to the topic, increasing the likelihood that users searching for climate-related datasets will find the dataset in question.

\begin{figure}[t]
    \small
    \centering
    \begin{tabular}{p{\textwidth}}
    \toprule
    \textbf{Search-Focused Description Prompt} \\
    \midrule
    \PromptFontSize \tt
    You are given a dataset about the topic \textcolor{purple}{\{D\_topic\}}, with the following initial description: \textcolor{purple}{\{D\_initial\_description\}}.\\
    \PromptFontSize \tt
    Please expand the description by including the exact topic. Additionally, add as many related concepts, synonyms, and relevant terms as possible based on the initial description and the topic.
    Unlike the initial description, which is focused on presentation and readability, the expanded description is intended to be indexed at backend of a dataset search engine to improve searchability.
    Therefore, focus less on readability and more on including all relevant terms related to the topic. Make sure to include any variations of the key terms and concepts that could help improve retrieval in search results.
    Please follow the structure of following example template:  \textcolor{purple}{\{Template\}}.\\
    \bottomrule
    \end{tabular}

    \vspace{.2cm}
    \input{sections/sfd-output-trimmed}
    \vspace{-.2cm}
    \caption{\revshep{Prompt and example of an abridged SFD.}} 
        \vspace{-.5cm}
    \label{tab:prompt_example_sfd}
\end{figure}


\autoref{tab:prompt_example_sfd} shows the prompt template used for generating an SFD which includes the dataset topic $D\_topic$ and an initial description $D\_initial\_description$ as inputs; the output structure defined by the SFD template.
%
The result is a description densely packed with terms relevant to the topic, which significantly improves the dataset's ranking in search results and makes it easier for users to discover the dataset when searching for related topics. A complete SFD example output is given in 
\subm{the extended version of this paper \cite{AutoDDG_arxiv}.}
\arxiv{Table~\ref{tab:example_sfd} in the Appendix.}

\subsubsection{Discussion: UFDs and SFDs}
While both UFD and SFD serve important roles in dataset description, they are optimized for different purposes. The UFD is crafted with the end user in mind, making the description easy to understand and presenting a well-rounded summary of the dataset. In contrast, the SFD is more technical and keyword-driven, focusing on enhancing search engine performance rather than prioritizing human readability.
The generation of both types of descriptions allows \SystemName to effectively serve both user-friendly and search-optimized needs.

There are different ways in which these descriptions can be generated. Here, we describe our initial approach that treats the SFD as an extension of the UFD.
In the SFD, semantic information about the dataset’s topic is reinforced to generate expanded keywords, concepts, and use cases, as illustrated in \autoref{tab:prompt_example_sfd}.
Indexing SFD descriptions in search engines improves dataset findability by incorporating richer, more relevant terms.
In future work, we plan to explore different strategies to combine the data-driven and LLM-derived information.

%% file: tables/example_sp.tex
%
    
    
%

\begin{figure}[t!]
    \small
    \centering
    
    \begin{tabular}{@{}p{0.48\linewidth}@{\hspace{0.04\linewidth}}p{0.48\linewidth}@{}}
    \toprule
    \textsf{\textbf{Example 1:} \textit{Year} column with values like 2018, 2020, 2023.} & \textsf{\textbf{Example 2:} \textit{Liabilities} column with values like 137790801, 43992755, 599895.} \\
    \begin{tcolorbox}[colback=black!2.5!white,colframe=black!85!black,boxrule=0.25mm,boxsep=4pt,left=0pt,right=0pt,top=0pt,bottom=0pt]
    \scriptsize
    \tt
    \begin{verbatim}
Temporal:
 - isTemporal: True
 - resolution: Year
Spatial:
 - isSpatial: False
 - resolution:
Entity Type: Temporal Entity
Domain-Specific Types: General
Function/Usage Context:
  Aggregation Key
    \end{verbatim}
    \vspace{-1.5em}
    \end{tcolorbox}
    &
    \begin{tcolorbox}[colback=black!2.5!white,colframe=black!85!black,boxrule=0.25mm,boxsep=4pt,left=0pt,right=0pt,top=0pt,bottom=0pt]
    \scriptsize
    \tt
    \begin{verbatim}
Temporal:
 - isTemporal: False
 - resolution:
Spatial:
 - isSpatial: False
 - resolution:
Entity Type: Monetary Value
Domain-Specific Types: Financial
Function/Usage Context:
  Measurement
    \end{verbatim}
    \vspace{-1.5em}
    \end{tcolorbox}
    \end{tabular}
    
    \vspace{-.4cm}
    \caption{\revshep{Examples of semantic information derived from datasets by the Semantic Profiler.}}
    \vspace{-.3cm}
    \label{fig:examples_semantic_profiler}
\end{figure}

%














%% file: sections/sfd-output-trimmed.tex
    \textsf{Example Abridged Output:}
    \vspace{-.5em}
    \begin{tcolorbox}[colback=black!2.5!white,colframe=black!85!black,boxrule=0.25mm,boxsep=4pt,left=0pt,right=0pt,top=0pt,bottom=0pt]
    \PromptFontSize
    \tt \raggedright
    
    \textcolor{teal}{\textit{Dataset Overview:}} This dataset contains wind speed and direction measurements {\color{gray} [... an overview based on the UFD.]}
    \\
    \textcolor{teal}{\textit{Related Topics:}}
    - Climate/Weather Patterns
    - Renewable Energy
    - Wind Energy
    - Meteorology
    ...
    \\
    \textcolor{teal}{\textit{Concepts and Synonyms:}}
    - Wind Speed/Velocity
    - Wind Direction
    - Average Wind Speed
    ...
    \\
    \textcolor{teal}{\textit{Applications and Use Cases:}}
    - Analysis of wind patterns for renewable energy projects
    - Understanding climate trends and predicting wind behavior
    ...
    \\
    \textcolor{teal}{\textit{Additional Context:}}
    - This dataset can be used to address questions such as "What are the typical wind patterns in a given region?" or "How does climate change affect wind behavior?"
    - It can be integrated with other datasets, such as climate models or energy consumption data, to provide a more comprehensive understanding of the relationship between wind patterns and energy generation.
    {\color{gray} [\emph{additional information omitted}] }
    \end{tcolorbox}

%% file: sections/Sec4_Benchmark.tex
\section{Evaluation and New Benchmarks}

We propose \revgen{the use of} multiple strategies to assess both the \textit{intrinsic} quality of descriptions and their impact on dataset retrieval. Next, we describe the evaluation strategies and metrics we use 
as well as new benchmarks that we have derived from existing resources to support the evaluation of dataset descriptions. 

    \vspace{-.15cm}
\subsection{Evaluating Description Quality}
\label{sec:solution_evaluation}

We use three complementary strategies to evaluate description quality: (1) \textit{dataset retrieval evaluation}, which assesses the impact of descriptions on search effectiveness, (2) \textit{reference-based evaluation}, which compares generated descriptions to existing ones using natural language generation metrics, and (3) \textit{reference-free evaluation}, which leverages LLMs to assess descriptions without reference texts. The following subsections detail each approach.


\myparagraph{Dataset Retrieval Evaluation}
Evaluating the impact of generated descriptions on dataset retrieval performance is critical, as keyword-based search remains the primary application for dataset search engines. This evaluation assesses explicitly the ability of description generation models to enhance dataset discoverability by improving search relevance and ranking--arguably the most critical metric for practical use cases.
To measure retrieval performance, we use the Normalized Discounted Cumulative Gain (NDCG@k) metric~\cite{jarvelin2002ndcg}, a widely-used standard for ranking evaluation. NDCG@k measures the effectiveness of a ranking by comparing the ideal ranking of relevant items to the system's actual ranking up to position k. It accounts for both the relevance of the datasets and their positions in the result list, assigning higher scores to datasets that appear earlier. 

In our evaluation, we integrate the generated descriptions into a keyword-based search engine and perform search queries representative of typical user behavior. By calculating NDCG@k scores for search results with and without the generated descriptions, we can quantify the improvement in dataset discoverability attributable to our method.
This retrieval-based evaluation provides a direct measure of description effectiveness in supporting dataset search engines. Unlike intrinsic evaluations, which focus on the quality of the descriptions themselves, this extrinsic evaluation measures their practical impact on improving search outcomes.

\myparagraph{Reference-Based Evaluation}
\revised{For datasets that already have existing descriptions, we can compare generated descriptions with the original ones. This allows us to assess whether our automated approach produces descriptions that have similar quality to those written by humans. This is particularly important for data portals seeking to enhance or update their metadata, ensuring that automated descriptions meet user expectations.}


\noindent \textsf{METEOR}: The METEOR score~\cite{banerjee2005meteor} combines precision, recall, synonymy, and stemming for a more flexible comparison between the generated and reference descriptions. 

\noindent \textsf{ROUGE}: The ROUGE score~\cite{lin2004rouge} focuses on recall by measuring the overlap of n-grams, longest common subsequences, and skip-bigrams between the generated and reference texts.

\noindent \textsf{BERTScore}: BERTScore \cite{zhang2019bertscore} uses pre-trained language models, such as BERT, to measure the similarity of embeddings between the generated text and the reference text. Unlike ROUGE and METEOR, it evaluates semantic similarity by comparing contextualized word representations, providing an assessment of whether the generated description conveys the same information as the reference. \revgen{In this work, we report the F1 variant of BERTScore.} 

These metrics provide a structured, quantitative evaluation of how well the generated description matches the reference in terms of wording, content, and meaning.

\myparagraph{Reference-Free Evaluation}
For datasets without reference descriptions, reference-free evaluation is essential. In this scenario, we leverage large language models (LLMs) as judges to evaluate the quality of the generated descriptions. Instead of relying on direct comparison with reference texts, LLM-based evaluation assesses attributes such as coherence, relevance, clarity, and coverage of key dataset features~\cite{liu2023geval, gao2024llmEval}. We also conduct expert evaluation on a sample of descriptions to
validate their faithfulness to the dataset and the reliability of LLM
evaluation results. 


\noindent \textbf{LLM-Based Evaluations}: 
\revised{We first adopt \textit{\textsf{Pointwise}} evaluation }by following the prompt design from G-Eval \cite{liu2023geval}, incorporating task introduction, evaluation criteria, and evaluation steps. Additionally, we include example evaluations to guide the LLM in rating dataset descriptions. 
%
\revised{In this setting, we evaluate each candidate item individually based on the following criteria: \textit{completeness}, \textit{conciseness}, and \textit{readability}. }
For instance, compared with search-focused descriptions (SFD), user-focused descriptions (UFD) are expected to score higher on conciseness and readability but lower on completeness.
\revised{In addition, we use \textit{\textsf{Pairwise}} evaluation in which an LLM is presented with two candidate descriptions and must select the superior one \citep{liusie2024llm,cao2024compassjudger}.
}
\revised{For each dataset, we uniformly sample ten unordered pairs of descriptions generated by distinct methods. Each pair is judged twice—once for each presentation order—to mitigate positional bias. Similarly, the LLM evaluates the candidates along three dimensions: \textit{completeness}, \textit{conciseness}, and \textit{readability}. We report the \emph{win rate}, defined as the number of victories divided by the total number of pairwise comparisons in which the method participates.} 
%
%
%
%

However, it is important to acknowledge that LLM-based evaluations may introduce biases inherent in the models' training data. The models might favor certain styles or content, especially when evaluating outputs generated by the same model \cite{panickssery2024llm_favor}. 
To mitigate this issue, we utilize cross-evaluation --- for example, we use Llama to evaluate descriptions generated by GPT (and vice versa). By employing different models for generation and evaluation, we reduce the likelihood of shared biases influencing the assessment.

\noindent \revthree{\textbf{Expert Evaluation}:
To validate the LLM-based metrics, three annotators evaluated a stratified sample of 48 datasets (12 from ECIR-DDG, 
36 from NTCIR-DDG) across four description types: AutoDDG variants (SFD-GPT 
and UFD-GPT), original descriptions, and LLM-GPT (our strongest baseline). Annotators assessed four dimensions on a 1--10 scale: \textit{completeness}, 
\textit{conciseness}, \textit{readability} (matching automated metrics), 
and \textit{faithfulness}---the accuracy with which descriptions represent actual dataset content without hallucinations or factual errors, including the dataset's variables, structure, and statistics, etc. The evaluation guidelines are given in 
\subm{the extended version of this paper \cite{AutoDDG_arxiv}.}
\arxiv{Table~\ref{tab:llm_eval_comp_conc_read} in the Appendix.}
We assess inter-annotator agreement (IAA) and LLM judge-human alignment by comparing GPT and Llama evaluations against normalized human consensus using Pearson's $r$ on z-score normalized ratings.}

Reference-free evaluation complements traditional metrics by incorporating semantic and contextual assessments, which are crucial for datasets with no or low-quality descriptions.

\subsection{New Dataset Retrieval Benchmarks}
\label{sec:dataset_benchmark}

\revthree{
Since our primary goal is to generate descriptions that improve dataset retrieval, 
we need benchmark collections that include not only tabular datasets but also queries and human-labeled results.}
Moreover, to capture realistic and challenging real-world scenarios, we must include tables that contain a large number of rows and columns.
Therefore, we consider the following criteria to select table search benchmarks: (1)~support for keyword-based queries; (2)~inclusion of CSV datasets collected from Open Data Portals; and (3)~availability of query relevance data, typically represented as triples of (CSV dataset, keyword query, relevance score).
While there are several table search benchmarks~\cite{lin2022acordar1, chen2024acordar2, hulsebos2023gittables, chen2021wtr, zhang2018semsearch, zhang2021semantic, leventidis2024STSD, ji2024target}, only two meet these criteria:
ECIR~\cite{chen2020ecir} and  NTCIR~\cite{kato2021ntcir}. 
\revthree{Below, we describe how we adapted them to evaluate the effectiveness of description generation for keyword-based dataset retrieval. The new benchmarks are summarized in Table~\ref{tab:benchmark_stats}.}

\begin{table*}
    \centering
    \caption{Statistics of the ECIR-DDG and NTCIR-DDG benchmarks: number of queries, the average, minimum, and maximum number of relevant tables per query (Rel. Tabs/Query), total tables per query (Tabs/Query), total tables per benchmark (Tabs/Bench), rows per table (Rows/Tab) and columns per table (Cols/Tab). These statistics highlight the diversity and scale of the datasets in the benchmarks.
    }
    \vspace{-.5em}
    \resizebox{\textwidth}{!}{
    \begin{tabular}{c|c|ccc|ccc|c|ccc|ccc}
        \toprule
        \bf Benchmark & \bf Query & \multicolumn{3}{c|}{\bf Rel. Tabs/Query} & \multicolumn{3}{c|}{\bf Tabs/Query} & \bf Tabs/Bench & \multicolumn{3}{c|}{\bf Rows/Tab} & \multicolumn{3}{c}{\bf Cols/Tab} \\
        \hline
        & & Avg & Min & Max & Avg & Min & Max & & Avg & Min & Max & Avg & Min & Max \\
        \hline
        \bf ECIR-DDG & 120 & 12 & 8 & 16 & 169 & 76 & 549 & 1015 & 17977 & 4 & 183539 & 22 & 4 & 337 \\
        \bf NTCIR-DDG & 32 & 10 & 5 & 22 & 19 & 10 & 42 & 615 & 78630 & 3 & 2717008 & 14 & 2 & 74\\
        \bottomrule
    \end{tabular}
    }
    \label{tab:benchmark_stats}
    \vspace{-0.25cm}
\end{table*}

\myparagraph{ECIR-DDG Benchmark}
\revthree{ECIR contains 2,417 tabular datasets published by the U.S. Federal Government
in \textsf{data.gov}~\cite{datagov} and consists of six tasks, covering various domains such as public health, economic data, and environmental statistics, each of which describes an information need and the associated relevant datasets along with relevance scores. 
For each task, 20 distinct keyword queries were created using Mechanical Turk, and relevance judgments for \textit{task-dataset pairs} were obtained through crowdsourcing.} 
\revthree{
%
To validate the benchmark's  \textit{query-dataset pairs}, for each query $q$, we  randomly sampled datasets $D_{rel}$ labeled as relevant and datasets $D_{irrel}$  labeled as irrelevant for $q$. 
}


We manually reviewed the dataset snippets—including the title, description, column names, and a sample of rows—to determine how many datasets in $D_{rel}$ were truly relevant to the query $q$ (true positives) and how many in $D_{irrel}$ were truly irrelevant to $q$ (true negatives). 
Manual validation revealed a true-positive rate of 9.87\%, with 23 out of 233 dataset-query pairs ($D_{rel}, q$) identified as valid. The true negative rate was 98.5\%, with 65 out of 66 dataset-query pairs ($D_{irrel}, q$) confirmed as irrelevant.

\revthree{Given the low true-positive rate, we used Google Dataset Search (GDS) to construct a new collection of relevant datasets for the queries. We issued the keyword queries to GDS 
and manually verified the retrieved datasets for relevance. The collected datasets, along with their original titles and descriptions, considered relevant to the benchmark. We retain the original irrelevant candidates without modification. }

%

\myparagraph{NTCIR-DDG Benchmark}
\revthree{NTCIR was built as a community effort carried out by six research groups.
They collected data from Data.gov (U.S. Government's open data)~\cite{datagov} and e-Stat (Japan Government's open data)~\cite{estat}, comprising documents and datasets across domains such as climate, population, economy, and transportation. Queries were constructed from real user questions found in community Q\&A forums and refined through crowdsourcing. They include geographical, temporal, and numerical terms, reflecting common information needs in dataset search. 
We focus on the English component of NTCIR which encompasses
datasets in different formats (e.g., CSV, PDF, JSON, XML and RDF). Given that we aim to evaluate descriptions of tabular data, we use only CSV datasets.
%
%
To ensure sufficient data for meaningful evaluation, we selected queries with at least five relevant datasets and at least 10 total datasets (relevant and irrelevant combined). This filtering process resulted in a benchmark with 32 queries and a total of 615 datasets (Table~\ref{tab:benchmark_stats}).}
\revthree{We manually validated a sample of the queries and found the relevance assessment to be accurate.}


%% file: sections/Sec5_Experiment.tex
\section{Experimental Evaluation}
\label{sec:experiment}

\subsection{Experiment Overview}
To evaluate our dataset description generation framework, we conduct experiments following the evaluation strategy introduced in Section~\ref{sec:solution_evaluation} using multiple baselines and description models.

%
We consider the following models for constructing descriptions:\\
(1) \textbf{Original}: the dataset descriptions provided with the datasets. 
(2) \textbf{Header+Sample}: a simple method that concatenates column headers with sample values from each column.

\noindent Additionally, we evaluate two pre-trained language models by providing them with the Header and Sample: \\
(3) \textbf{MVP}~\cite{tang2023mvp}: the Multi-task Supervised Pre-training for Natural Language Generation (MVP) model leverages multi-task training for improved text generation, including data-to-text generation.\\
(4) \textbf{ReasTAP} \cite{zhao2022reastap}: a table pre-training model that incorporates reasoning skills to enhance table comprehension, including table-to-text generation.\\
\revised{
(5) \textbf{Pneuma} \cite{pneuma@sigmod2025}: an end-to-end retrieval-augmented generation (RAG) system designed to support tabular data discovery using natural language queries; \revshep{we use its \texttt{Table Summarizer}, which generates schema summaries via LLM and row summaries by mechanically pairing randomly sampled values with column headers.}
}
\\
\revgen{(6) \textbf{LLM-GPT/Llama}: an LLM-based baseline (GPT or Llama) that generates descriptions from 5 randomly sampled rows, using the same sampling strategy as AutoDDG.} 

\revone{We use the HuggingFace models available
for MVP~\cite{mvpimpl}  and  ReasTAP~\cite{reastapimpl}. Given the limitations of these models with respect to window size, we provide as input samples of rows. For Pneuma, we use its authors' implementation~\cite{pneumaimpl}.} 

Finally, we evaluate our proposed AutoDDG framework, which uses large language models (LLMs) to generate dataset descriptions automatically. 
We selected two cost-effective LLMs for our experiments: \texttt{GPT-4o-mini}\footnote{\url{https://platform.openai.com/docs/models}} and \texttt{LLaMA-3.1-8B-Instruct}\footnote{\url{https://deepinfra.com/meta-llama/Meta-Llama-3.1-8B-Instruct}}, which we refer to as "GPT" and "Llama" respectively in method names. We instantiate AutoDDG with both models:\\
(7) \textbf{\SystemName-UFD-GPT/Llama}: generates user-focused descriptions (UFDs) optimized for readability and clarity while maintaining relevance for search.\\
(8) \textbf{\SystemName-SFD-GPT/Llama}: generates search-focused descriptions (SFDs) designed to improve dataset retrieval in keyword-based search engines.


%
%
%
%


\subsection{Retrieval Performance}
\label{sec:retrieval-performance}

To evaluate the effectiveness of dataset descriptions in search tasks, we measure retrieval performance with \textbf{BM25}~\cite{robertson2009bm25} and \textbf{SPLADE}~\cite{formal2021splade}. 
BM25 focuses on lexical matching, ranking documents based on statistics derived from term frequency and inverse document frequency.  
In contrast, SPLADE retrieves documents using vector search over sparse text representations, which are derived by expanding query terms with semantically related words.
For example, it may expand `lunar' to include `moon', thereby capturing broader context.

%
%
%


We experimented with both the ECIR-DDG and NTCIR-DDG benchmarks. 
%
We created indices for both BM25 and SPLADE using the dataset descriptions generated by the models.
For BM25, we created an inverted index, in which terms from each dataset description were tokenized and stored along with their term frequencies and document frequencies.\footnote{We used Okapi BM25 from this library: \url{https://github.com/dorianbrown/rank_bm25}} During evaluation, keyword-based queries were issued against this index, and BM25 ranked the dataset descriptions by calculating their relevance scores.
For SPLADE, we created a sparse vector representation of each dataset description.\footnote{We used SPLADE with FastEmbed: \url{https://qdrant.github.io/fastembed/examples/SPLADE_with_FastEmbed/}} 
During evaluation, keyword-based queries were processed, and the dataset descriptions were ranked by computing cosine similarity between the query's sparse representation and each document's sparse embedding.

\input{tables/eval_table_4_3_bm25}

\myparagraph{BM25 Results}
%
The results in Table~\ref{tab:ndcg_bm25} show that \SystemName-SFD methods consistently outperform UFD across both ECIR-DDG and NTCIR-DDG benchmarks (for a given LLM), confirming the effectiveness of SFDs for retrieval. By tailoring descriptions to maximize keyword matching with query topics, SFDs improve search relevance.
\revthree{The GPT-4o-mini model leads to higher NDCG values than Llama on the ECIR-DDG benchmark, and Llama shows better performance on the NTCIR-DDG benchmark. This suggests that different LLMs exhibit varying strengths depending on the characteristics of the dataset and query topics.}
%

The performance of UFD is comparable to that of SFD. However, as expected, the focus on user readability limits its effectiveness in purely search-oriented metrics. 
Traditional baselines such as MVP and ReasTAP perform significantly worse than both UFD and SFD, highlighting their limitations in generating effective descriptions for retrieval tasks. Interestingly, original descriptions sometimes outperform these baselines, underscoring the importance of domain-specific context and the limitations of these data-to-text approaches in keyword-based retrieval.

\begin{figure}
	\centering
	\begin{subfigure}{0.35\columnwidth} 
		\centering
		\includegraphics[width=1.0\linewidth]{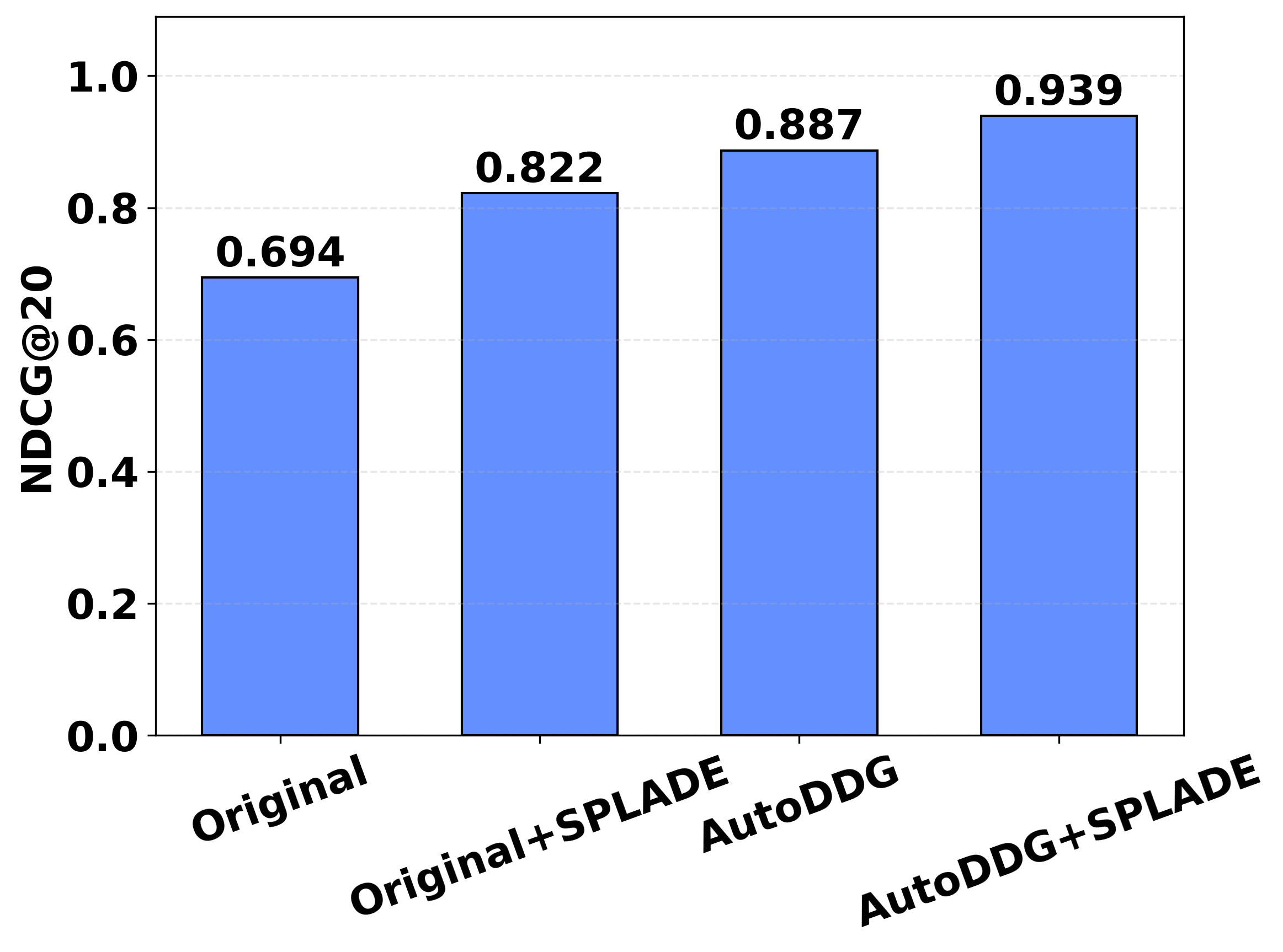}
		\caption{ECIR-DDG Benchmark}
        \label{fig:ablation_noSTA}
	\end{subfigure}
        \begin{subfigure}{0.35\columnwidth} 
		\centering
		\includegraphics[width=1.0\linewidth]{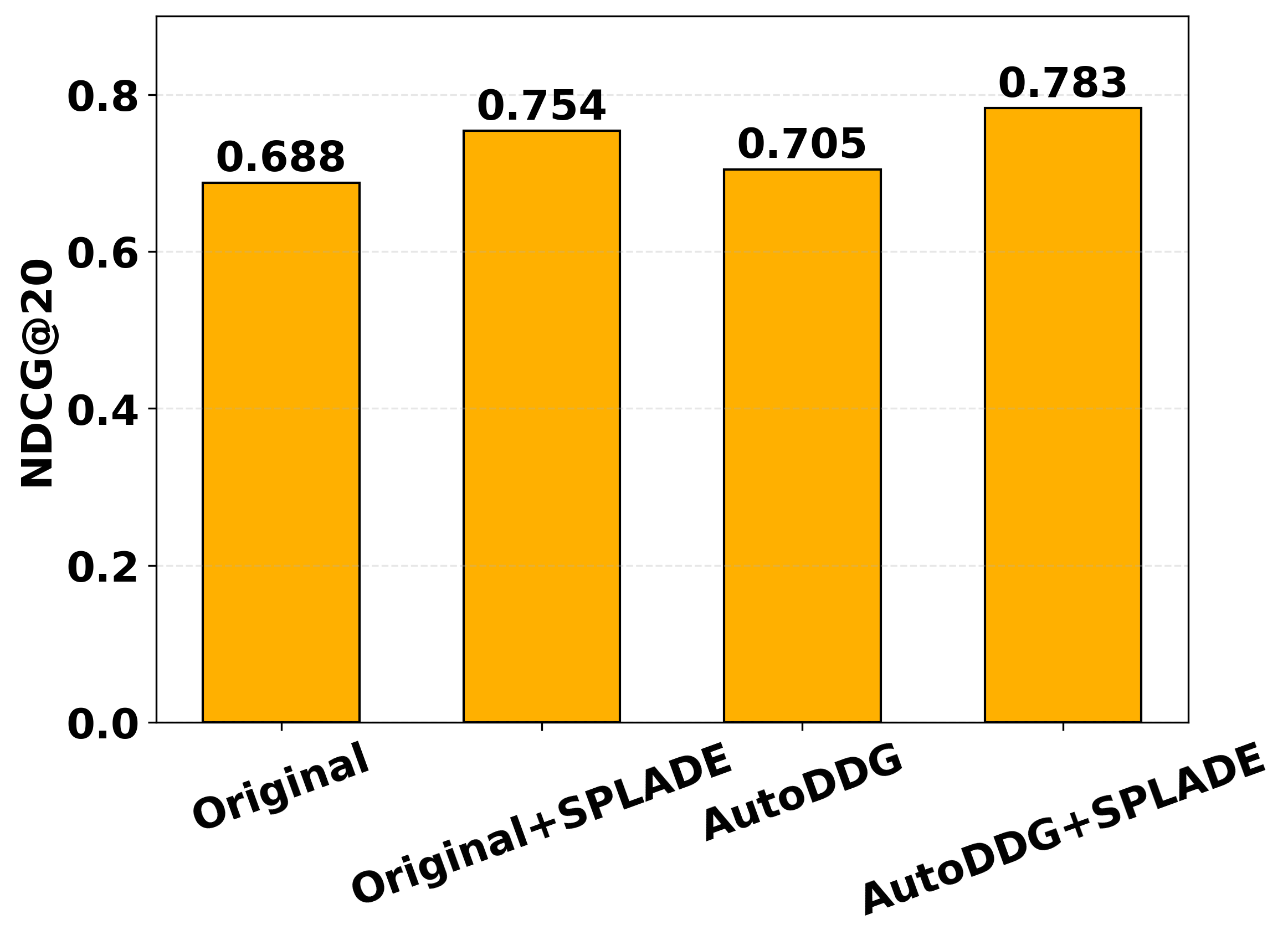}
		\caption{NTCIR-DDG Benchmark}
        \label{fig:ablation_ufd}
	\end{subfigure}
	\caption{\revshep{Comparison of NDCG@20 scores for Original, Original+SPLADE, AutoDDG, and AutoDDG+SPLADE on ECIR-DDG and NTCIR-DDG benchmarks shows improvements from combining AutoDDG and SPLADE.}}
	\label{fig:summary_retrieval}
    \vspace{-0.5cm}
\end{figure}

\myparagraph{SPLADE Results}
\revised{
\autoref{fig:summary_retrieval} compares retrieval performance using the original and \SystemName's SFD descriptions with BM25 and SPLADE. 
The results show that SFD descriptions consistently outperform the original descriptions when using BM25's lexical matching. 
Similarly, we observe that SPLADE's semantic term expansion in the vector space yields significant gains over BM25 on the original description.
This is expected, since both \SystemName and SPLADE use different approaches to expand the information contained in the original descriptions.
However, the best results are achieved by combining SPLADE with AutoDDG descriptions. These results suggest that the approaches are complementary and that AutoDDG's SFD descriptions provide additional helpful information not present in the original description.
}
%
%
\subm{A complete table of NDCG scores with SPLADE is available in the extended version of this paper~\cite{AutoDDG_arxiv}.}
\arxiv{A complete table of NDCG scores with SPLADE is available in the Appendix~\Cref{tab:experiment_splade}.}
\subsection{Evaluation of Dataset Description Quality}
\label{sec:description-quality}


We follow the methodology introduced in Section~\ref{sec:solution_evaluation} to evaluate the quality of the generated dataset descriptions using both reference-based and reference-free methods. We: 1) assess alignment with ground-truth descriptions using different methods---METEOR, ROUGE, and BERTScore, which capture phrase similarity, lexical overlap, and semantic consistency; 
2) evaluate the descriptions' intrinsic quality using LLM-based scoring for completeness, conciseness, readability--these metrics assess how well the descriptions convey relevant information while maintaining clarity and accuracy; 3)  \revgen{carry out a user evaluation in which experts assess descriptions to validate both their faithfulness to source datasets and the reliability of LLM-based evaluation.}

This multi-pronged approach to evaluation offers insights into how well the descriptions meet human readability standards and perform in automated search and retrieval tasks. This is useful for users seeking to publish new datasets, as these evaluations can predict the performance of the generated descriptions for both front-end user interaction and back-end dataset retrieval. 





\begin{table*}[t!]
\centering
\caption{\textit{Pointwise} evaluation of dataset description quality on reference-free metrics and \textit{Pairwise} evaluation across all models for both the ECIR-DDG and NTCIR-DDG benchmarks. Scores are reported as GPT/Llama pairs. Completeness (Comp), Conciseness (Conc), and Readability (Read) are scaled from 0 to 10, while Pairwise is scaled from 0 to 1. Higher values indicate better performance. Bold values denote the highest scores for each metric, and underlined values denote the second-highest.}
\label{tab:eval_reference_all}
\resizebox{\textwidth}{!}{
\begin{tabular}{l|cccc|cccc}
\toprule
 & \multicolumn{4}{c|}{\bf ECIR-DDG} & \multicolumn{4}{c}{\bf NTCIR-DDG} \\
\bf Model & \bf Comp (G/L) & \bf Conc (G/L) & \bf Read (G/L) & \bf Pairwise (G/L) & \bf Comp (G/L) & \bf Conc (G/L) & \bf Read (G/L) & \bf Pairwise (G/L) \\
\hline
Original & 4.07/5.02 & 6.83/7.35 & 6.04/7.14 & \revgen{0.35/0.31} & 4.46/4.81 & 7.04/6.14 & 6.17/6.37 & \revgen{0.34/0.39} \\
\hline
H+S & 3.50/3.56 & 5.39/4.00 & 4.36/4.15 & \revgen{0.18/0.26} & 3.24/3.57 & 5.01/3.92 & 3.98/4.57 & \revgen{0.24/0.30} \\
MVP & 1.52/1.57 & 3.24/3.25 & 2.41/2.84 & \revgen{0.05/0.09} & 1.29/1.51 & 2.57/2.90 & 1.96/2.43 & \revgen{0.03/0.11} \\
ReasTAP & 1.96/1.91 & 4.38/5.81 & 3.33/4.85 & \revgen{0.03/0.07} & 1.69/1.83 & 3.63/4.81 & 2.54/3.98 & \revgen{0.04/0.06} \\
Pneuma & 5.25/7.09 & 5.65/4.78 & 5.19/5.81 & \revgen{0.28/0.46} & 5.26/6.64 & 5.89/5.27 & 5.47/6.32 & \revgen{0.28/0.48} \\
\revgen{LLM-GPT} & \revgen{6.40/6.75} & \revgen{8.50/\textbf{8.00}} & \revgen{8.17/8.08} & \revgen{0.50/0.45} & \revgen{6.40/6.54} & \revgen{8.30/7.87} & \revgen{8.15/8.03} & \revgen{0.48/0.46} \\
\revgen{LLM-Llama} & \revgen{6.17/6.50} & \revgen{8.26/7.92} & \revgen{8.04/8.03} & \revgen{0.52/0.45} & \revgen{6.03/6.17} & \revgen{8.23/7.82} & \revgen{8.00/7.91} & \revgen{0.49/0.42} \\
\hline
UFD-GPT & 8.19/8.57 & \bf 8.97/8.00 & \bf 8.99/8.99 & \revgen{0.73/\underline{0.72}} & 8.17/8.58 & \bf 8.49/8.00 & \bf 8.89/8.99 & \revgen{0.77/\underline{0.68}} \\
UFD-Llama & 7.72/8.43 & \underline{8.51}/\underline{7.86} & \underline{8.63}/\underline{8.50} & \revgen{0.69/0.68} & 7.59/8.06 & \underline{8.33}/\underline{7.88} & \underline{8.42}/\underline{8.65} & \revgen{0.60/0.60} \\
SFD-GPT & \bf 8.96/8.98 & 7.35/5.97 & 7.90/7.00 & \revgen{\textbf{0.86}/0.71} & \bf 9.00/9.00 & 8.00/6.31 & 9.00/8.16 & \revgen{\underline{0.87}/0.70} \\
SFD-Llama & \underline{8.95}/\underline{8.89} & 6.24/5.56 & 7.19/6.79 & \revgen{\underline{0.77}/\textbf{0.78}} & \underline{8.99}/\underline{9.00} & 7.96/6.01 & 8.94/8.04 & \revgen{\textbf{0.79}/\textbf{0.70}} \\
\bottomrule
\end{tabular}
}
\vspace{-1em}
\end{table*}

\myparagraph{Reference-Based Evaluation}
%
The trends observed in the reference-based evaluation (Table~\ref{tab:eval_reference_based}) are consistent with those from other evaluations. The AutoDDG-UFD models perform exceptionally well on \textit{METEOR}, indicating improved semantic alignment with reference descriptions. This suggests that UFDs, designed with human-centered presentation in mind, better capture the intent and meaning of the original data description. In contrast, AutoDDG-SFD achieves superior scores in \textit{ROUGE}, signaling a higher overlap with the reference text's exact terms and phrases. This aligns with the typical goal of search-focused descriptions: ensuring high precision in matching key data points. \revgen{The AutoDDG methods achieve slightly lower BERTScores than LLM-only baselines, likely because BERTScore favors embedding similarity to reference phrasing. In contrast, our approach prioritizes factual grounding, generating semantically accurate but linguistically diverse descriptions, as evidenced by competitive METEOR and ROUGE scores.}

\input{tables/eval_table_4_3_reference}

\myparagraph{Reference-Free Evaluation}
The results of this evaluation, summarized in Table~\ref{tab:eval_reference_all}, show that AutoDDG descriptions consistently outperform other methods. AutoDDG-UFD-GPT excels in \textit{Conciseness} and \textit{Readability}, confirming the expectation that user-focused descriptions (UFDs) prioritize clarity and readability. This reinforces our design decision of creating user-centered descriptions. 
Conversely, AutoDDG-SFD-GPT ranks highest in \textit{Completeness}, as expected from search-focused descriptions (SFDs) that prioritize comprehensive data coverage, even at the expense of readability or conciseness.
\revgen{
The LLM-only methods achieve competitive \textit{Conciseness} scores by generating shorter descriptions without profiling, but their significantly lower \textit{Completeness} underscores their incomplete knowledge of the data.}
%
%
Furthermore, the \textit{Pairwise} comparison included in Table~\ref{tab:eval_reference_all} also demonstrates that AutoDDG outperforms other methods.
\revthree{Despite numerical differences, GPT and Llama evaluations show consistent trends across all reference-free metrics.} 
%
%
%
%

\begin{table}[t!]
\caption{\revgen{Human evaluation of description quality across methods. Scores represent mean $\pm$ standard deviation on a 1--10 scale, computed from z-score normalized ratings across three annotators ($n=192$ per method). Inter-annotator agreement (IAA) is measured by Pearson's $r$ on normalized scores.}}
\label{tab:human_eval_results}

\centering
\begin{small}
\revgen{\begin{tabular}{lcccc|c}
\toprule
\textbf{Model} & \textbf{Comp.} & \textbf{Conc.} & \textbf{Read.} & \textbf{Faith.} & \textbf{Average} \\
\hline
Original & $4.33 \pm 1.68$ & $6.45 \pm 1.75$ & $6.35 \pm 1.81$ & $5.47 \pm 1.65$ & $5.65 \pm 0.85$ \\
LLM-GPT & $6.95 \pm 0.76$ & \underline{$8.73 \pm 0.55$} & \underline{$8.83 \pm 0.58$} & $7.59 \pm 1.19$ & $8.03 \pm 0.79$ \\
UFD-GPT & \underline{$8.67 \pm 0.50$} & \boldmath$\mathbf{8.75 \pm 0.70}$ & \boldmath$\mathbf{8.99 \pm 0.61}$ & \underline{$9.00 \pm 0.89$} & \boldmath$\mathbf{8.85 \pm 0.14}$ \\
SFD-GPT & \boldmath$\mathbf{9.37 \pm 0.45}$ & $7.40 \pm 0.55$ & $7.61 \pm 0.58$ & \boldmath$\mathbf{9.09 \pm 0.82}$ & \underline{$8.37 \pm 0.87$} \\
\hline
IAA (avg $r$) & 0.792 & 0.477 & 0.494 & 0.593 & 0.719 \\
\bottomrule
\end{tabular}%
}
\end{small}

\end{table}

\myparagraph{\revgen{User Evaluation}}
\revgen{To validate LLM-based evaluation, we conduct a user evaluation on a stratified sample of 48 datasets (192 descriptions total). Three users assessed the faithfulness of the generated descriptions by comparing them with the original datasets. We normalize user scores to remove individual annotator bias prior to analysis. 
Table~\ref{tab:human_eval_results} shows that AutoDDG descriptions have substantially higher faithfulness than the ones derived by LLM-GPT. This indicates 
that the integration of data profilers and topic generators effectively reduces hallucinations and factual errors---a critical advantage over pure LLM generation. Original descriptions achieve only
$5.47 \pm 1.65$ for faithfulness, showing high variance and indicating inconsistent quality.
Both AutoDDG methods also excel in completeness and overall quality, which aligns with the LLM-as-judge results. 
Annotator comments reveal that original descriptions provide only metadata and external links rather than content descriptions, achieving \textit{"conciseness at the cost of completeness"}. LLM-only descriptions frequently contain \textit{factual errors including wrong temporal ranges, incorrect statistics, and unsupported inferences}, which AutoDDG methods largely eliminate. Users identified occasional errors in the AutoDDG descriptions, which we traced to bugs in the algorithmic profiler—for example, misidentifications of temporal coverage or units of measure. Nonetheless, the descriptions were consistent with the information output by the profiler.}

\begin{figure}
  \centering
  \includegraphics[width=.7\linewidth]{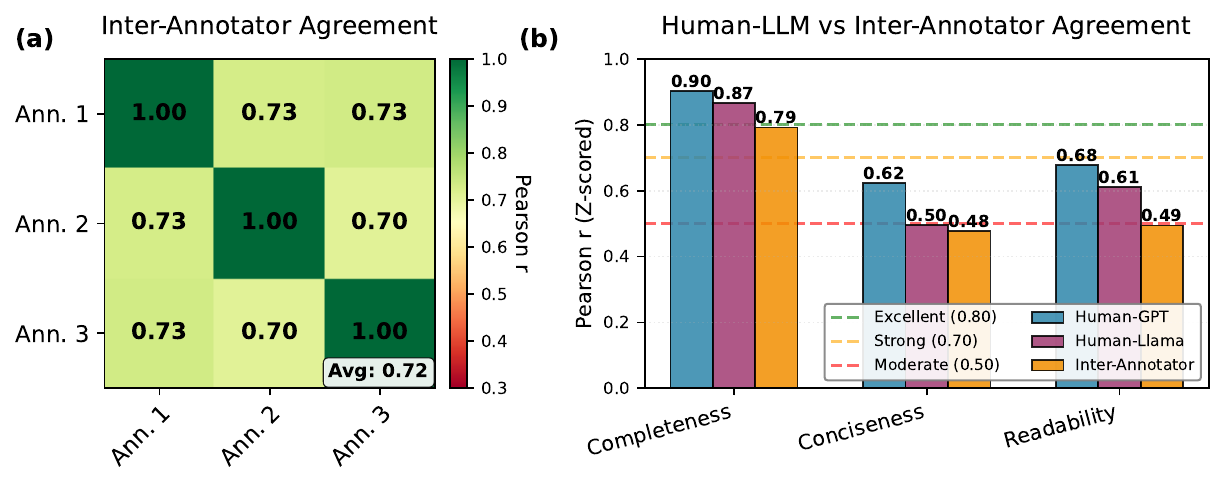}
  \caption{\revgen{\textbf{(a)} Inter-annotator agreement on overall assessment (avg $r=0.72$). \textbf{(b)} Human-LLM vs inter-annotator agreement across three metrics. All correlations use Pearson's $r$ on z-score normalized ratings ($n=192$, $p<0.001$).}}
  \label{fig:human_eval_vs_llm}
  \vspace{-0.5cm}
\end{figure}

\revgen{ 
Figure~\ref{fig:human_eval_vs_llm}  shows that there is a high correlation between human annotators and the scores derived by the LLMs, confirming the reliability of the automated LLM evaluation. 
%
Panel (a) shows substantial inter-annotator agreement (r=0.72) on overall 
assessment, computed by averaging the four scores for each dataset 
before measuring agreement between annotators. Panel (b) reveals strong LLM-human agreement on objective completeness metrics. Inter-annotator agreement on conciseness and readability is lower, reflecting the subjective nature of these dimensions. By analyzing the comments from the annotators, we can see that they considered different stylistic criteria: \textit{one prioritized logical flow and appropriate detail levels for dataset complexity, while another focused on redundancy across sections and representative examples.} This variation is expected for subjective writing assessments where individual preferences naturally differ. Notably, LLM evaluators achieve better agreement with mean human ratings (GPT: $r=0.62$, Llama: $r=0.5$), indicating that automated evaluation effectively aggregates diverse human judgments on these subjective dimensions. Overall, 
LLM-as-judge achieves human-level reliability for assessing description quality. See Appendix for detailed per-metric breakdowns.}

\myparagraph{Summary}
Our experimental evaluation confirms our hypothesis: UFDs score higher in conciseness, readability, and semantic coherence (as seen in \textit{BERTScore}), while SFDs perform better in completeness and exact overlap with reference descriptions. These insights suggest that, although UFDs are more suitable for user assessment and semantic clarity, SFDs are better suited to search-related tasks that benefit from more detailed descriptions.
Therefore, using both descriptions in a dataset search engine improves both the user experience and the quality of search results.

\revshep{\subsection{Cost Analysis}}
\revshep{We evaluate the LLM-related computational cost and scalability of the AutoDDG pipeline across a sample of 283 datasets from the NTCIR and ECIR benchmarks. We compare our approach against all baselines except the heuristic-based H+S, whose runtime is negligible.}
\revshep{To ensure a fair comparison, all LLM-based methods (AutoDDG, Pneuma, and LLM-GPT) utilize \texttt{gpt-4o-mini} as the underlying backbone. 
For conciseness, we report results for the AutoDDG UFD-GPT configuration; for SFD, the cost is incremental, requiring one additional LLM call using the UFD output.}


\newcommand{\autoddgmt}{\emph{AutoDDG-MT}\xspace}
\newcommand{\autoddggp}{\emph{AutoDDG-GP}\xspace}
\newcommand{\autoddgseq}{\emph{AutoDDG-Seq}\xspace}

\revshep{
To mitigate the cost and runtime overheads inherent in making multiple LLM API calls for the Semantic Profiling stage (Algorithm~\ref{alg:semantic_type_analyzer}), we implemented two optimization strategies:\\
--\autoddgmt (multi-threaded) applies Algorithm~\ref{alg:semantic_type_analyzer} concurrently by
issuing distinct API calls for each column, utilizing a pool of 64 worker threads to improve throughput. \\
--\autoddggp (group prompting) utilizes a batching strategy where multiple columns are grouped into a single prompt. This enables the sharing of system instructions across columns within a single context window, thereby reducing the total number of input tokens. \\
We refer to the sequential strategy as \autoddgseq.
} 

\begin{figure*}[!t]
  \centering
  \includegraphics[width=0.95\linewidth]{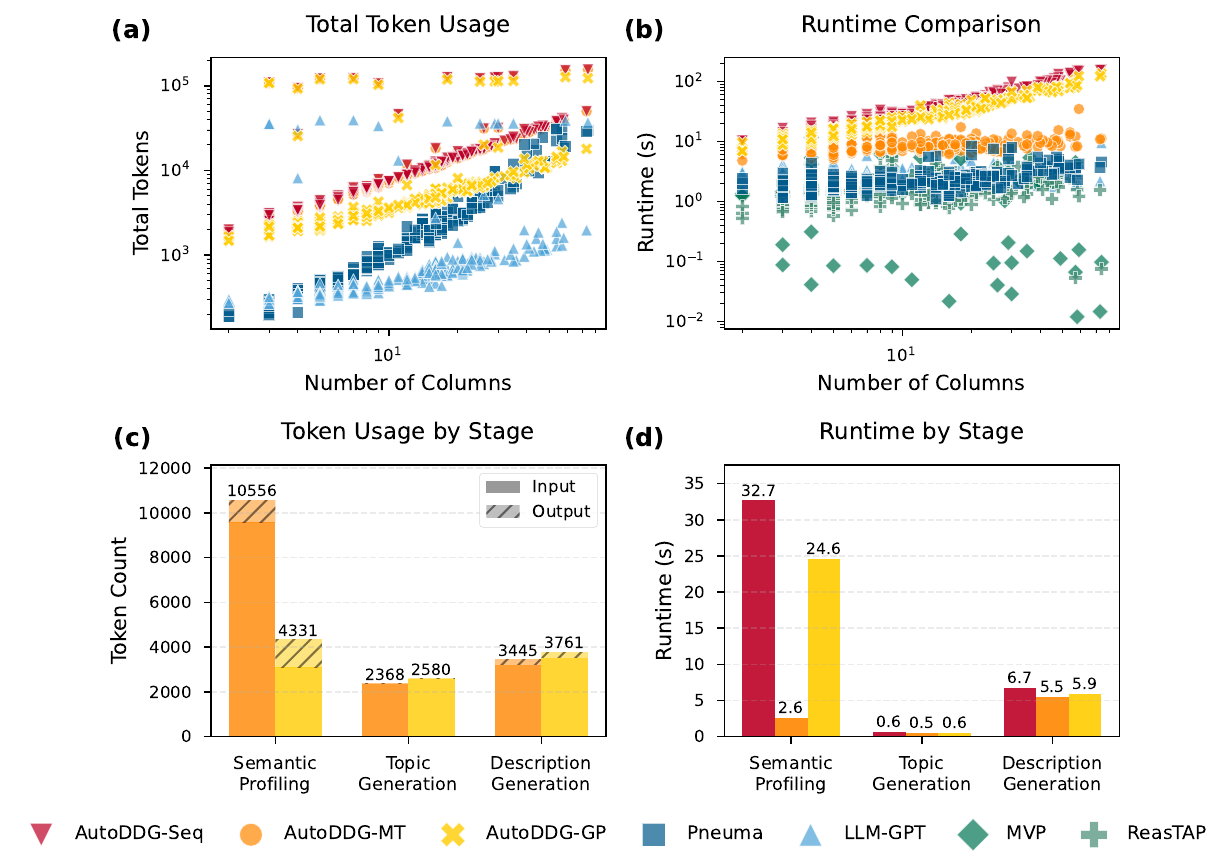}
  \caption{
  \revshep{Cost and scalability analysis across 283 datasets. (a-b) runtime and LLM token usage scaling relative to dataset width (log-log scale)  (token usage is shown for LLM-based methods). AutoDDG maintains predictable scalability despite higher overhead compared to baselines. (c-d) Stage-wise cost breakdown for AutoDDG. The comparison between multi-threading (orange) and group prompting (yellow) reveals a key trade-off: group prompting reduces input tokens but increases latency due to serialization bottlenecks--its runtime is similar to that of the AutoDDG sequential variant (red).}
  }
  \label{fig:cost_analysis}
  \vspace{-0.3cm}
\end{figure*}

\noindent \revshep{
\myparagraph{Scalability: Runtime and Token Usage}  
We assessed the scalability of the different approaches with respect to dataset width--varying number of columns.
For AutoDDG,  the total runtime includes semantic profiling, topic generation, and description generation. To ensure a fair comparison, we also ran the Pneuma baseline using multiple threads. The results are shown in  Figure~\ref{fig:cost_analysis}.}

\revshep{
The concurrent execution of API calls results in a substantial performance improvement: for the maximum number of columns, \autoddgmt is about 10 times faster than \autoddgseq.
}
\revshep{
\autoddgmt has predictable scaling with dataset width. While its runtimes are longer than those of the baselines, they are of the same order of magnitude. On average, \autoddgmt takes 8.63 sec per dataset, compared to 2.40 sec for Pneuma, 2.76 sec for LLM-GPT, 1.93 sec for MVP, and 2.01 sec for ReasTAP.}
\revshep{To test enterprise-level scalability, we ran AutoDDG-MT over a table with 337 columns--it completed the task in 17.16 seconds.}
%


\revshep{
The token usage for the different LLM-based methods is shown in Figure~\ref{fig:cost_analysis}(b). Since the token usage is identical for \autoddgseq and \autoddgmt, we present the results for the former. As expected, the number of tokens used by AutoDDG grows with dataset width.
The new group prompting strategy (\autoddggp) is effective at reducing token usage: it reduces the average number of tokens per dataset from 16,370 to 10,674 and significantly dampens the growth rate for wide tables.
%
Note that AutoDDG uses more tokens than the LLM-only baseline (which issues a single prompt per dataset) and Pneuma--while Pneuma, like AutoDDG, issues one call per column, it  does not execute the token-heavy description-generation step.}

\revshep{In Figure~\ref{fig:cost_analysis}(b), we also observe coincident token-count outliers for AutoDDG and the LLM-GPT baseline, which are absent for Pneuma. While all three approaches use an LLM, their input strategies differ: AutoDDG and LLM-GPT use sampled row values to construct prompts, resulting in token counts that increase substantially when datasets contain very long values in their columns. Since Pneuma does not include value samples during LLM-powered schema narration, it does not incur token costs for those samples. Importantly, Figure 10(a) confirms that these input-token surges do not translate into runtime outliers, demonstrating that \autoddgmt remains efficient even when processing long textual values.}

\revshep{To further assess the scalability of the different methods, we experimented with a very wide table containing 337 columns. 
For this dataset, Pneuma consumed 1,202,309 tokens, whereas AutoDDG-MT required 218,760 tokens. This difference arises from their respective architectural designs: Pneuma incorporates the entire schema as context for each column description to maintain global coherence, whereas AutoDDG processes columns independently during semantic profiling. 
%
This shows that engineering systems to use LLMs efficiently is challenging, as many trade-offs must be considered. It also suggests that AutoDDG could serve as a complementary tool for Pneuma’s table-summarization stage, particularly for handling datasets with many columns.}

\revshep{
We find that the modest overhead in runtime and token consumption for AutoDDG is justified by the significant performance gains in both retrieval effectiveness (Table~\ref{tab:ndcg_bm25}) and generated description quality (Tables~\ref{tab:eval_reference_all}, \ref{tab:eval_reference_based}, and \ref{tab:human_eval_results}). 
%
%
} 

\noindent \revshep{\myparagraph{Optimizing LLM Usage} 
%
We further analyze the trade-offs between the two optimization strategies. As  Figure~\ref{fig:cost_analysis}(d) shows, group prompting (yellow) is highly effective at reducing cost. Sharing system instructions reduces the semantic-profiling input token count by approximately 60\%. However, Figure~\ref{fig:cost_analysis}(c) reveals a critical trade-off between token and runtime efficiency: multi-threaded single-column prompting (orange) achieves far superior performance ($\sim$2.6s) compared to group prompting ($\sim$24.6s).
This performance gap stems from the auto-regressive nature of LLM decoding: even when columns are grouped, the semantic profiling is performed sequentially for each column  
within a single response window (intra-batch serialization). \autoddgmt 
eliminates this serialization bottleneck. In practice, we can use multi-threading for latency-sensitive applications, while group prompting remains a viable cost-saving alternative for resource-constrained environments where high concurrency is not feasible.}

\noindent \revshep{\myparagraph{AutoDDG: Total Cost} The description generation stage dominates the remaining computational costs, averaging 3,445 tokens and 5.5 seconds per dataset. The total 
cost per dataset is approximately \$0.003 (using \texttt{gpt-4o-mini}), with Group Prompting further reducing this to $\approx$\$0.0023, confirming AutoDDG's applicability for large
dataset collections.}

\begin{figure}[t]
  \centering
  \includegraphics[width=0.45\linewidth]{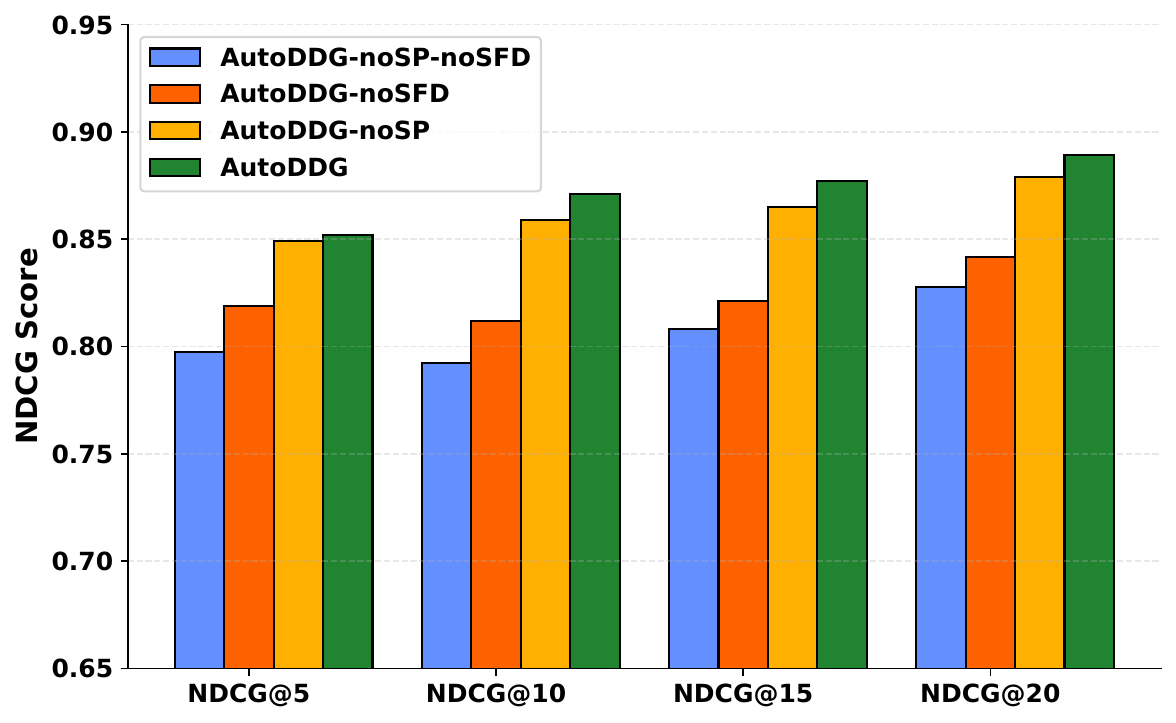}
    \includegraphics[width=0.45\linewidth]{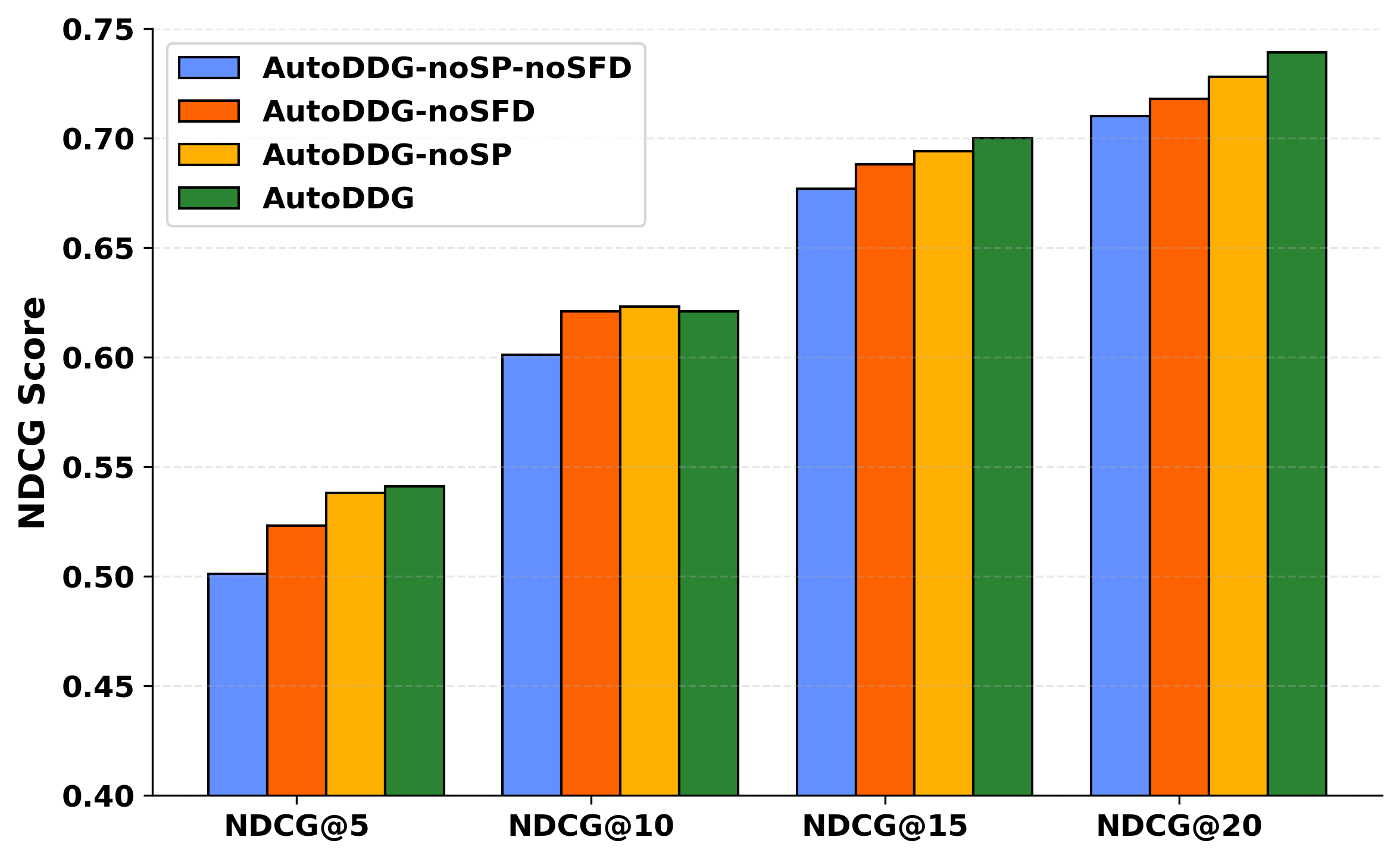}
  \caption{\revone{Comparison of NDCG scores across different settings of \SystemName (noSP-noSFD, noSFD, noSP and full configuration) on BM25 for the ECIR-DDG (left) and NTCIR-DDG (right) benchmarks, showing the impact of enabling semantic profile (SP) and search-focused description (SFD) on retrieval performance.}}
  \label{fig:ablation_different_settings}
  \vspace{-.5cm}
\end{figure}

\subsection{Ablation Study}

To assess the contributions of each module in \SystemName, we conduct an ablation study by selectively disabling key components and analyzing their impact on retrieval performance. This allows us to isolate the effects of the Semantic Profiler (SP)  and the Search-Focused Description (SFD).  
We compare the following variations:  
(1) \textbf{\SystemName-noSP-noSFD}: Generates descriptions without SP and SFD, relying only on basic dataset samples and statistics.  
(2) \textbf{\SystemName-noSFD}: Includes SP but omits SFD, enhancing descriptions with semantic insights while not optimizing for search retrieval.  
\revone{
(3) \textbf{\SystemName-noSP}: Includes SFD but omits SP, test the configuration optimizing for search retrieval without SP.
}
(4) \textbf{\SystemName}: Includes both SP and SFD, leveraging semantic insights while optimizing descriptions for search retrieval.

\myparagraph{Impact of Different Modules in \SystemName}  
Figure~\ref{fig:ablation_different_settings} summarizes the results, 
measured by NDCG scores.  
The inclusion of the semantic profile (SP) module (\SystemName-noSFD) improves retrieval compared to the baseline (\SystemName-noSP-noSFD).  
\revone{
Adding the search-focused description (SFD) module alone (\SystemName-noSP) provides further gains. The \SystemName with all components has the best performance, reinforcing its role in enhancing search relevance.
}

\begin{figure}[t]
  \centering
  \includegraphics[width=0.55\linewidth]{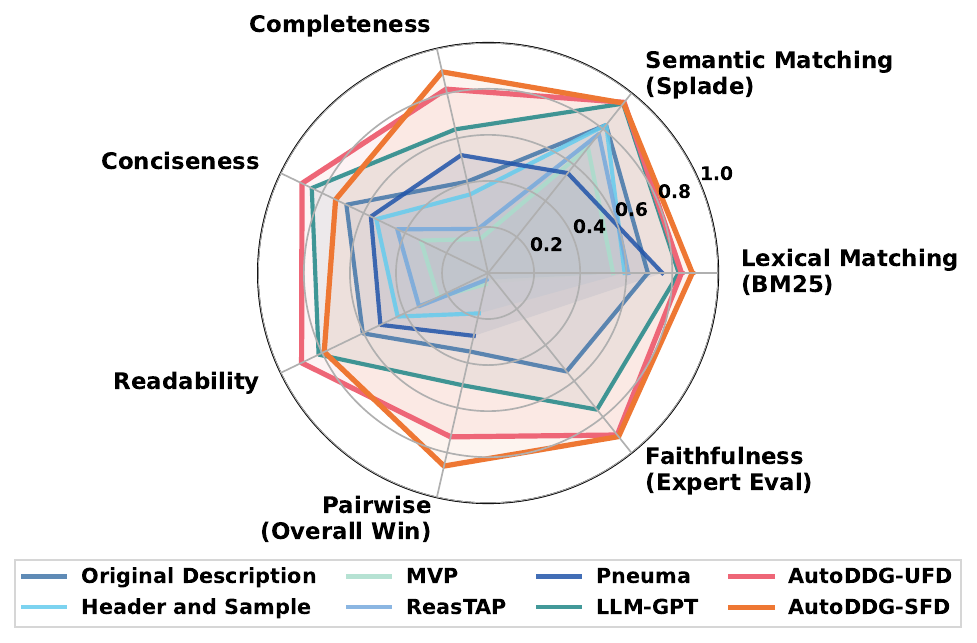}
  \caption{Radar chart comparing the performance of various dataset description methods across evaluation metrics.
  }
  \label{fig:summary_radar}
  \vspace{-0.5cm}
\end{figure}

\subsection{Overview of Experimental Results}
\revone{The radar chart in Figure~\ref{fig:summary_radar} provides a comparative overview, illustrating how different methods perform across these dimensions.}
The results show that UFD excels in readability and conciseness, making it more suitable for user-facing applications, while SFD improves completeness and retrieval relevance, increasing dataset findability. By combining these strengths, AutoDDG achieves a balanced and superior performance compared to baseline methods (Original, Header+Sample, MVP, and ReasTAP), demonstrating its effectiveness in generating high-quality descriptions for both human interpretation and machine-based search.

%% file: tables/eval_table_4_3_bm25.tex
\begin{table*}[!t]
    \centering
    \caption{BM25. Comparison of NDCG scores across different description generation models on the ECIR-DDG and NTCIR-DDG benchmarks. Higher NDCG scores indicate better alignment between the generated descriptions and the relevance of retrieved results for keyword-based search. Bold values represent the highest scores for each metric, while underlined values indicate the second highest.}
    \resizebox{\textwidth}{!}{
    \begin{tabular}{l|cccc|cccc}
        \toprule
         & \multicolumn{4}{c|}{\bf ECIR-DDG} & \multicolumn{4}{c}{\bf NTCIR-DDG} \\
         \bf Model & \bf NDCG@5 & \bf NDCG@10 & \bf NDCG@15 & \bf NDCG@20 & \bf NDCG@5 & \bf NDCG@10 & \bf NDCG@15 & \bf NDCG@20 \\
        \hline
         Original & 0.666 & 0.658 & 0.668 & 0.694 & 0.439 & 0.575 & 0.662 & 0.688\\
         \hline
         Header+Sample & 0.589 & 0.573 & 0.575 & 0.593 & 0.323 & 0.451 & 0.553 & 0.590\\
         MVP & 0.557 & 0.538 & 0.533 & 0.543 & 0.356 & 0.472 & 0.575 & 0.617\\
         ReasTAP & 0.638 & 0.598 & 0.599 & 0.609 & 0.406 & 0.485 & 0.583 & 0.627\\
         Pneuma & 0.652 & 0.695 & 0.725 & 0.757 & 0.423 & 0.561 & 0.651 & 0.682\\
         \revgen{LLM-GPT} & \revgen{0.794} & \revgen{0.800} & \revgen{0.803} & \revgen{0.828} & \revgen{0.415} & \revgen{0.536} & \revgen{0.615} & \revgen{0.651}\\
         \revgen{LLM-Llama} & \revgen{0.774} & \revgen{0.783} & \revgen{0.784} & \revgen{0.802} & \revgen{0.422} & \revgen{0.559} & \revgen{0.638} & \revgen{0.674}\\
         \hline
         AutoDDG-UFD-GPT & \underline{0.800} & \underline{0.804} & \underline{0.815} & \underline{0.840} & 0.415 & 0.553 & 0.645 & 0.682\\
         AutoDDG-UFD-Llama & 0.760 & 0.762 & 0.777 & 0.801 & 0.425 & 0.565 & 0.652 & 0.679 \\
         AutoDDG-SFD-GPT & \bf 0.849 & \bf 0.856 & \bf 0.867 & \bf 0.887 & \bf 0.458 & \underline{0.585} &\bf 0.672 & \bf 0.705 \\
         AutoDDG-SFD-Llama & 0.766 & 0.785 & 0.806 & 0.828 & \underline{0.456} & \bf 0.594 & \underline{0.668} & \underline{0.696}\\
      \bottomrule
    \end{tabular}
    \label{tab:ndcg_bm25}
    }
    \vspace{-0.5em}
\end{table*}

%% file: tables/eval_table_4_3_reference.tex
\begin{table*}[!t]
\centering
\caption{Evaluation of dataset description quality on reference-based metrics for both the ECIR-DDG and NTCIR-DDG benchmarks. Higher values indicate better performance. Bold values represent the highest scores for each metric, while underlined values indicate the second highest.
}
\begin{small}
\begin{tabular}{l|ccc|ccc}
\toprule
 & \multicolumn{3}{c|}{\bf ECIR-DDG} & \multicolumn{3}{c}{\bf NTCIR-DDG} \\
\bf Model & \bf METEOR & \bf ROUGE & \bf BERTScore & \bf METEOR & \bf ROUGE & \bf BERTScore \\
\hline
Original & - & - & - & - & - & - \\
\hline
Header+Sample & 2.28 & 3.61 & 75.17 & 2.41 & 4.25 & 75.46 \\
MVP & 1.55 & 1.83 & 78.85 & 1.58 & 2.60 & 78.85 \\
ReasTAP & 6.48 & 8.82 & 79.99 & 4.94 & 6.71 & 79.93 \\
Pneuma & 7.58 & 26.97 & 77.68 & 10.24 & 28.29 & 78.48 \\
\revgen{LLM-GPT} & \revgen{11.93} & \revgen{20.14} & \revgen{\underline{82.45}} & \revgen{12.60} & \revgen{22.28} & \revgen{\underline{82.89}} \\
\revgen{LLM-Llama} & \revgen{12.04} & \revgen{19.03} & \revgen{\bf 82.85} & \revgen{13.27} & \revgen{22.72} & \revgen{\bf 83.11} \\
\hline
AutoDDG-UFD-GPT & \underline{15.48} & 29.78 & 82.24 & \bf 15.90 & \underline{30.56} & 82.47 \\
AutoDDG-UFD-Llama & \bf 16.46 & 30.08 & 82.26 & \underline{15.64} & 27.80 & 82.60 \\
AutoDDG-SFD-GPT & 13.00 & \bf 34.50 & 80.07 & 14.49 & \bf 35.45 & 80.24 \\
AutoDDG-SFD-Llama & 12.27 & \underline{33.85} & 79.17 & 13.80 & 32.48 & 79.40 \\
\bottomrule
\end{tabular}
\label{tab:eval_reference_based}
\end{small}
\end{table*}

%% file: sections/Sec2_RelatedWork.tex
\section{Related Work}
\label{sec:related_work}

\myparagraph{Dataset Search}
\revshep{Studies on information-seeking behavior reveal significant challenges in dataset discovery. Koesten et al.~\cite{koesten2017trials} found that users often cannot find needed data and identified common search strategies and key dataset attributes in queries: category/topic, geographic region, data granularity (spatial and temporal resolution), and relevance indicators like summary statistics and data semantics. }
%
\revshep{\citet{papenmeier2021genuine} reported  that ``current systems fail to sufficiently support scientists in their data-seeking process'' and because the metadata associated with datasets does not cover information that is sought by scientists. This underscores the disconnect between users' information needs and the datasets they can actually find, and the importance of data descriptions that accurately characterize the data~\cite{chapman2020dataset}.}
\revshep{\citet{Sostek2024Discovering} confirmed these metadata shortcomings in Google Dataset Search and identified an additional challenge: inconsistency in dataset descriptions increases mental load during relevance assessment.}
%
We use the findings from these studies to guide the design of the data-driven and semantic summaries in \SystemName. 
By using the dataset's actual contents to generate summaries and augmenting them with semantic information, we can derive high-quality descriptions. While it may not be possible to derive completely uniform descriptions, given that datasets contain different information, an automated process can be applied to all datasets to enforce a common structure. 
For example, it is not possible to summarize the spatial extent of a dataset that lacks spatial data; however, for all spatial datasets, this information can be provided and represented in a similar manner.

\revshep{Discovery queries that go beyond keyword search have been proposed in which a dataset $D$ is the query, and the results include datasets that are related to $D$, e.g., can be joined or concatenated, or are correlated~\cite{lazo@icde2019,josie@sigmod2019,santos23,santos2021correlation}. These queries are orthogonal to and can be combined with keyword search.}
%

\myparagraph{Table Understanding and Representation}
\revshep{Many models have been developed to enhance table understanding and facilitate various downstream tasks by semantically representing table content \cite{hulsebos2019sherlock, wang2021tcn, zhang2019sato, deng2022turl, archetype@vldb2024, herzig2020tapas, chorus@vldb2024, korini2023columnGPT, liu2021tapex, suhara2022doduo, yin2020tabert, wang2021tuta, iida2021tabbie}. 
These areas include, but are not limited to, table question answering, column-type analysis, and table-type classification.
Recently, large language models (LLMs) have begun to play a significant role in various data-related tasks, including column type annotation (CTA)~\cite{korini2023columnGPT,chorus@vldb2024,archetype@vldb2024}
and table-class detection~\cite{chorus@vldb2024}.
%
While these approaches advance semantic understanding of tables, they do not directly address the problem of generating dataset descriptions.
Nonetheless, the information they derive can be used to enrich dataset descriptions. For example, \SystemName uses the strategy proposed in \cite{chorus@vldb2024} to obtain the dataset topic.
In future work, we will explore incorporating additional semantic information, such as column types~\cite{archetype@vldb2024}.}

\myparagraph{Text Generation from Tables}
Table-to-text generation models have attained significant attention due to their ability to transform structured data into coherent natural language text \cite{chen2019d2t_lm_switch, puduppully2019d2t_ncp, gong2020tablegpt, chen2020kgpt, liu2021d2t_augplan, su2021d2t_p2g, wang2021d2t_sketch, puduppully2021d2t_macro, zhao2022reastap, tang2023mvp, seo2024unveiling}. 
Among recent advancements, ReasTAP \cite{zhao2022reastap} introduces a novel table pre-training approach that enhances models' reasoning capabilities through synthetic question-answering examples and demonstrates notable performance gains in producing logically faithful text across various downstream tasks. Additionally, the Multi-task Supervised Pre-training for Natural Language Generation (MVP) model \cite{tang2023mvp} leverages multi-task supervised pre-training to excel in a broad range of natural language generation tasks, including knowledge-graph-to-text and data-to-text generation. 
However, these models are trained on datasets such as Rotowire \cite{wiseman2017rotowire}, WikiBio \cite{lebret2016wikibio}, and LogicNLG \cite{chen2020logicNLG}, which are designed to generate narrative text tailored to specific tasks, such as sports game summaries and biographical sentences.
While effective for generating text from structured data, these models lack the ability to: (1) create descriptions for keyword-based search and high-level overviews; (2) handle large and heterogeneous datasets with complex structures.
Moreover, they require training for a specific objective. In contrast, we aim to generate summaries that are sufficiently general to accommodate a wide range of datasets and address diverse information needs, without requiring specific training for each dataset.

\revised{Pneuma~\cite{pneuma@sigmod2025} is an end-to-end retrieval-augmented generation (RAG) system designed to support tabular data discovery using natural language queries. 
\revshep{While Pneuma and \SystemName aim to improve dataset discoverability, they have important technical
differences.
Pneuma leverages LLMs to narrate table schemas to extract attribute semantics and relies on row samples to represent the dataset's contents.}
In contrast, \SystemName applies both a data-driven and an LLM-powered profiler to obtain both semantic information and derive a global summary of the dataset contents. 
In Section~\ref{sec:experiment}, we show that AutoDDG has better retrieval performance than Pneuma, suggesting that Pneuma could benefit from integrating AutoDDG's approach for description generation.}

\myparagraph{\revtwo{LLM-based Metadata Extraction from Documents}}
\revtwo{
Automating metadata extraction from scholarly documents using LLMs is an emerging research area, though challenges remain regarding robustness and accuracy~\cite{watanabe2024capabilities, alyafeai2025mextract}.
Frameworks like MOLE~\cite{alyafeai2025mole} employ LLMs to extract and validate over 30 distinct attributes that describe datasets (e.g., license, volume, tasks) from full-text papers. 
Data Gatherer~\cite{marini2025data} focuses on extracting structured dataset references (identifiers and repository names) from open-access scientific publications available in databases such as PubMed Central.
In domain-specific applications, LLMs were utilized to complete missing metadata fields in physical sample records (e.g., catalog records or specimen records)~\cite{song2024metadata}. 
Effective strategies for this task include record summarization before classification, few-shot prompting, and integrating hierarchical domain taxonomy via bottom-up traversal.
While these works share AutoDDG's goal of improving resource findability, they differ in the types of resources and in the approaches to analyzing data and generating LLM context.
}

\myparagraph{LLMs as Judges}
\revised{LLMs have recently been explored as scalable, consistent evaluators of natural language generation (NLG) outputs, offering an alternative to costly and variable human assessments. These automated evaluations typically follow two paradigms: \emph{pointwise} scoring and \emph{pairwise} comparison \citep{li2024llms}. The pointwise approach assigns numerical ratings to
outputs based on dimensions such as fluency, relevance, and faithfulness. For example, \citet{chiang2023can} demonstrated that GPT-4’s Likert-scale ratings for story generation and adversarial text examples align closely with expert human judgments, and \citet{wang2023chatgpt} reported that ChatGPT’s pointwise evaluations on summarization, story generation, and data-to-text tasks achieve state-of-the-art correlations with human references. In contrast, pairwise evaluation requires the model to choose the better of two candidates.
This method provides more discriminative judgments~\citep{liusie2024llm} and has been confirmed to closely approximate human rankings in large-scale studies~\citep{bavaresco2024llms}.}
\revised{Our evaluation design is inspired by these prior works.}

%% file: sections/Sec6_Conclusion.tex
\vspace{-.3cm}
\section{Conclusions and Future Work}
\label{sec:conclusion}

Effective dataset descriptions are essential for improving findability and helping users assess dataset relevance. However, many datasets lack informative descriptions, limiting their discoverability and usability in search systems. In this paper, we introduce \SystemName, an end-to-end framework for automated dataset description generation that systematically addresses this challenge.
\SystemName combines data-driven profiling with LLM-powered semantic augmentation to generate high-quality descriptions that balance comprehensiveness, faithfulness, conciseness, and readability. 
We proposed a multi-pronged evaluation strategy for dataset descriptions that measures improvements in dataset retrieval, assesses alignment with existing descriptions, and employs LLM-based scoring for intrinsic quality evaluation. Recognizing the limitations of existing benchmarks, we introduced two new dataset search benchmarks, ECIR-DDG and NTCIR-DDG, to enable rigorous assessment.

Our experimental results demonstrate that AutoDDG significantly improves dataset retrieval performance, produces high-quality, accurate descriptions, and helps users better understand and assess datasets. Beyond its immediate impact on dataset search engines, our framework provides a scalable and systematic approach to metadata enhancement, benefiting data repositories, open data portals, and enterprise data lakes.

\myparagraph{Limitations and Future Directions}
While \SystemName demonstrates strong performance for tabular datasets, an important avenue for future work is to extend its applicability to a broader range of dataset types and content structures. In particular, we aim to explore multi-modal approaches that incorporate additional data modalities, such as images, to generate richer descriptions.

Our data-driven and semantic profilers were designed to extract information identified as necessary for dataset search~\cite{koesten2017trials,papenmeier2021genuine,Sostek2024Discovering}.  
\revshep{Expanding their capabilities through incorporating functional dependency analysis and domain-specific customization during table sampling could enhance their effectiveness by better capturing inter-column relationships and serving diverse research domains and use cases.}
%
\hide{ 
\revtwo{Currently, AutoDDG uses only the dataset content (schema + rows) as context. We deliberately excluded existing metadata to
evaluate AutoDDG’s ability to generate descriptions from data
alone. This design choice also demonstrates AutoDDG’s robustness:
It can generate quality descriptions even when existing metadata is
missing or inadequate, which is precisely the problem we aim to solve. In future work, we would like to explore the integration of additional context, including provenance, existing descriptions, or papers associated with the dataset.}
} 
%
\revshep{While our implementation relies on API-based lightweight models, transitioning to local GPU deployment would lead to lower cost and latency.
It would also enable optimizations such as batch inference, as used in~\citep{pneuma@sigmod2025}.}

A critical challenge in LLM-generated descriptions is hallucination, where the model may introduce incorrect information~\cite{huang2025survey}. Developing robust detection and mitigation strategies to ensure descriptions remain faithful to the dataset’s content and context is an important research direction.
Finally, while automatic description generation significantly improves dataset findability, the generated descriptions are necessarily incomplete. Exploring human-in-the-loop techniques, in which users can enrich (e.g., by adding provenance and information on data-collection methodology) or validate descriptions, offers an opportunity to further enhance accuracy, completeness, and trustworthiness, particularly in high-stakes domains. 

\section*{Acknowledgments}
This work was supported by NSF awards IIS-2106888 and OAC-2411221,  the DARPA
ASKEM program
Agreement No. HR0011262087, and the ARPA-H BDF program. The views, opinions, and findings expressed are those of the authors and should not be interpreted as representing the official views or policies of the DARPA, ARPA-H, the U.S. Government, or NSF.

%% file: sections/Sec7_Appendix.tex
\setcounter{topnumber}{3}
\setcounter{bottomnumber}{3}
\setcounter{totalnumber}{4}
\renewcommand{\topfraction}{1}
\renewcommand{\bottomfraction}{1}
\renewcommand{\textfraction}{0.1}

\pagebreak

\appendix
\clearpage

\section{Additional Experimental Results}
\input{tables/experiment_splade}

\newpage
\section{Prompts and Examples of Descriptions}
\input{tables/example_sfd}

\input{tables/prompt_sp_template}


\input{tables/prompt_semantic_profile}

\input{tables/template_sfd}

\input{tables/prompt_llm_eval}


%% file: tables/experiment_splade.tex
\begin{table*}[h]
    \centering
    \caption{SPLADE: Comparison of NDCG scores across different description generation models on the ECIR-DDG and NTCIR-DDG benchmarks. Higher NDCG scores indicate better alignment between the generated descriptions and the relevance of retrieved results for keyword-based search. Bold values represent the highest scores for each metric, while underlined values indicate the second highest. 
    }
    \resizebox{\textwidth}{!}{
    \begin{tabular}{l|cccc|cccc}
        \toprule
         & \multicolumn{4}{c|}{\bf ECIR-DDG} & \multicolumn{4}{c}{\bf NTCIR-DDG} \\
         \bf Model & \bf NDCG@5 & \bf NDCG@10 & \bf NDCG@15 & \bf NDCG@20 & \bf NDCG@5 & \bf NDCG@10 & \bf NDCG@15 & \bf NDCG@20 \\
        \hline
         Original & 0.833 & 0.826 & 0.820 & 0.822 & 0.614 & 0.698 & 0.734 & 0.754\\
         \hline
         Header+Sample & 0.808 & 0.794 & 0.802 & 0.823 & 0.445 & 0.561 & 0.628 & 0.665\\
         MVP & 0.722 & 0.706 & 0.687 & 0.696 & 0.427 & 0.524 & 0.606 & 0.649\\
         ReasTAP & 0.799 & 0.768 & 0.758 & 0.772 & 0.526 & 0.607 & 0.659 & 0.695\\
         Pneuma & 0.480 & 0.523 & 0.534 & 0.555 & 0.425 & 0.539 & 0.605 & 0.640\\
        LLM-GPT & 0.871 & \bf 0.922 & \underline{0.936} & 0.941 & 0.557 & 0.662 & 0.704 & 0.727\\
        LLM-Llama & 0.863 & 0.918 & 0.931 & 0.937 & 0.446 & 0.551 & 0.614 & 0.661\\
         \hline
         AutoDDG-UFD-GPT & \bf 0.902 & \underline{0.919} & \bf 0.940 & \bf 0.948 & \bf 0.646 & 0.695 & 0.754 & 0.775\\
         AutoDDG-UFD-Llama & 0.890 & 0.910 & 0.930 & 0.941 & 0.643 & \bf 0.716 & \bf 0.759 & \bf 0.788\\
         AutoDDG-SFD-GPT & \underline{0.898} & 0.916 & 0.935 & \underline{0.944} & 0.635 & 0.689 & 0.747 & 0.768\\
         AutoDDG-SFD-Llama & 0.885 & 0.910 & 0.926 & 0.939 & \underline{0.643} & \underline{0.715} & \underline{0.754} & \underline{0.783}\\
      \bottomrule
    \end{tabular}
    }
    \label{tab:experiment_splade}
\end{table*}

%% file: tables/example_sfd.tex
\begin{table*}[h!]
    \centering
   \resizebox{.75\textwidth}{!}{    
    \begin{tabular}{p{\textwidth}}
    \toprule
    \textbf{Search Focused Description (SFD) Example} \\
    \midrule
    \textcolor{teal}{\textbf{Dataset Overview:}} \\
    - The dataset provides comprehensive financial information pertaining to various health insurance companies, specifically focusing on different types of insurers, including Health Maintenance Organizations (HMO), Managed Care Health (MCH), and Acciden \& Health (A\&H) insurance providers. It includes pivotal financial attributes of these companies such as company name, year of data collection, total assets, liabilities, and premiums written. Spanning from 2013 to 2023, the dataset reflects a breadth of financial metrics, indicating the fiscal condition of these insurers and presents premium values categorized in monetary formats. This dataset aims to furnish insights into the financial landscape of the health insurance sector.
    \\\\
    \textcolor{teal}{\textbf{Key Themes or Topics:}} \\
    - Health Insurance Finance \\
    - Health Insurance Types (HMO, MCH, A\&H) \\
    - Financial Analysis in Insurance \\
    - Risk Assessment in Health Insurance \\
    - Insurance Premium Structures \\
    - Temporal Changes in Health Insurance \\
    \\
    \textcolor{teal}{\textbf{Applications and Use Cases:}} \\
    - Supports analysis for health insurance professionals and executives in evaluating financial health. \\
    - Assists policymakers in understanding the economic implications of health insurance. \\
    - Enables researchers to investigate trends and outcomes related to health insurance provisions. \\
    - Useful for actuarial studies and financial modeling pertaining to health insurance. \\
    - Facilitates comparative studies between different types of health insurance organizations. \\
    \\
    \textcolor{teal}{\textbf{Concepts and Synonyms:}} \\
    - Health Insurance/Medical Insurance \\
    - Financial Data/Financial Metrics \\
    - Insurance Providers/Insurers \\
    - Premium Income/Premiums Written \\
    - Liability Assessment/Financial Liabilities \\
    - Asset Valuation/Total Assets \\
    - Risk Management/Insurance Risk \\
    - Health Economics/Economic Aspects of Health Insurance \\
    - HMO/Managed Care Organizations \\
    \\
    \textcolor{teal}{\textbf{Keywords and Themes:}} \\
    - Health insurance dataset \\
    - Financial health of insurers \\
    - Insurance premiums \\
    - Assets and liabilities in insurance \\
    - Temporal financial data \\
    - Insurance sector analysis \\
    - HMO, MCH, A\&H \\
    - Company financial attributes \\
    - Health insurance trends \\
    \\
    \textcolor{teal}{\textbf{Additional Context:}} \\
    - The dataset is relevant for addressing current challenges and inquiries in the health insurance sector, especially regarding financial performance metrics during economic fluctuations. \\
    - Its integration with health policy research and economic studies can provide a deeper understanding of how financial factors influence healthcare access and coverage. \\
    - It can aid in developing frameworks for better risk management and investment strategies within the health insurance market. \\
    \bottomrule
    \end{tabular}
    }
    \caption{Search Focused Description (SFD) Example for a dataset on health insurance financials.}
    \label{tab:example_sfd}
\end{table*}

%% file: tables/prompt_sp_template.tex
\begin{table*}
    \small
    \centering
    \begin{tabular}{|p{3cm}|p{11cm}|}
    \hline
    \textbf{Category} & \textbf{Details} \\
    \hline
    \textbf{Temporal} & 
    \begin{itemize}
        \item \textbf{isTemporal}: Does this column contain temporal information? Yes or No.
        \item \textbf{resolution}: If Yes, specify the resolution (Year, Month, Day, Hour, etc.).
    \end{itemize} \\
    \hline
    \textbf{Spatial} & 
    \begin{itemize}
        \item \textbf{isSpatial}: Does this column contain spatial information? Yes or No.
        \item \textbf{resolution}: If Yes, specify the resolution (Country, State, City, Coordinates, etc.).
    \end{itemize} \\
    \hline
    \textbf{Entity Type} & What kind of entity does the column describe? (e.g., Person, Location, Organization, Product). \\
    \hline
    \textbf{Domain-Specific Types} & What domain is this column from? (e.g., Financial, Healthcare, E-commerce, Climate, Demographic). \\
    \hline
    \textbf{Function/Usage Context} & How might the data be used? (e.g., Aggregation Key, Ranking/Scoring, Interaction Data, Measurement). \\
    \hline
    \end{tabular}
    \caption{Template for classifying column data into semantic types.}
    \label{tab:template}
\end{table*}

%% file: tables/prompt_semantic_profile.tex
\begin{table*}
    \small
    \centering
    \begin{tabular}{p{14cm}}
    \toprule
    \textbf{Dataset Semantic Profiler Prompt} \\
    \midrule
    \textcolor{teal}{\textbf{Instruction:}} \\
    You are a dataset semantic analyzer. Based on the column name and sample values, classify the column into multiple semantic types. \\

    \textcolor{teal}{\textbf{Categories:}} \\
    Please group the semantic types under the following categories: \\
    - \textbf{Temporal} \\
    - \textbf{Spatial} \\
    - \textbf{Entity Type} \\
    - \textbf{Data Format} \\
    - \textbf{Domain-Specific Types} \\
    - \textbf{Function/Usage Context} \\

    \textcolor{teal}{\textbf{Template Reference:}} \\
    Following is the template: \texttt{\{TEMPLATE\}} \\

    \textcolor{teal}{\textbf{Rules:}} \\
    \begin{enumerate}
        \item The output must be a valid JSON object that can be directly loaded by \texttt{json.loads}. Example response is \texttt{\{RESPONSE\_EXAMPLE\}}.
        \item All keys from the template must be present in the response.
        \item All keys and string values must be enclosed in \textbf{double quotes}.
        \item There must be no trailing commas.
        \item Use \textbf{booleans (true/false)} and numbers without quotes.
        \item Do not include any additional information or context in the response.
        \item If you are unsure about a specific category, you can leave it as an empty string.
    \end{enumerate} \\

    \textcolor{teal}{\textbf{Dynamic Parameters:}} \\
    - Column name: \texttt{\{column\_name\}} \\
    - Sample values: \texttt{\{sample\_values\}} \\

    \bottomrule
    \end{tabular}
    \caption{Prompt for the dataset semantic profiler.}
    \label{tab:prompt_semantic_profile}
\end{table*}

%% file: tables/template_sfd.tex
\begin{table*}
    \centering
    \resizebox{.75\textwidth}{!}{    
    \begin{tabular}{p{14cm}}
    \toprule
    \textbf{Search-Focused Description Template} \\
    \midrule
    \textcolor{teal}{\textbf{Dataset Overview:}} \\
    - Please keep the exact initial description of the dataset as shown at the beginning of the prompt. \\
    \\
    \textcolor{teal}{\textbf{Key Themes or Topics:}} \\
    - Central focus on a broad area of interest (e.g., urban planning, socio-economic factors, environmental analysis). \\
    - Data spans multiple subtopics or related areas that contribute to a holistic understanding of the primary theme. \\
    \textbf{Example:} \\
    - theme1/topic1 \\
    - theme2/topic2 \\
    - theme3/topic3 \\
    \\
    \textcolor{teal}{\textbf{Applications and Use Cases:}} \\
    - Facilitates analysis for professionals, policymakers, researchers, or stakeholders. \\
    - Useful for specific applications, such as planning, engineering, policy formulation, or statistical modeling. \\
    - Enables insights into patterns, trends, and relationships relevant to the domain. \\
    \textbf{Example:} \\
    - application1/usecase1 \\
    - application2/usecase2 \\
    - application3/usecase3 \\
    \\
    \textcolor{teal}{\textbf{Concepts and Synonyms:}} \\
    - Includes related concepts, terms, and variations to ensure comprehensive coverage of the topic. \\
    - Synonyms and alternative phrases improve searchability and retrieval effectiveness. \\
    \textbf{Example:} \\
    - concept1/synonym1 \\
    - concept2/synonym2 \\
    - concept3/synonym3 \\
    \\
    \textcolor{teal}{\textbf{Keywords and Themes:}} \\
    - Lists relevant keywords and themes for indexing, categorization, and enhancing discoverability. \\
    - Keywords reflect the dataset's content, scope, and relevance to the domain. \\
    \textbf{Example:} \\
    - keyword1 \\
    - keyword2 \\
    - keyword3 \\
    \\
    \textcolor{teal}{\textbf{Additional Context:}} \\
    - Highlights the dataset's relevance to specific challenges or questions in the domain. \\
    - May emphasize its value for interdisciplinary applications or integration with related datasets. \\
    \textbf{Example:} \\
    - context1 \\
    - context2 \\
    - context3 \\
    \bottomrule
    \end{tabular}
    }
    \caption{Dataset description template outlining key themes, applications, concepts, keywords, and contextual relevance.}
    \label{tab:template_sfd}
\end{table*}

%% file: tables/prompt_llm_eval.tex
\begin{table*}
    \scriptsize
    \centering
    \begin{tabular}{p{\textwidth}}
    \toprule
    \textbf{Dataset Description Evaluation Guidelines} \\
    \midrule
    \textcolor{teal}{\textbf{Task:}} 
    You will be given one tabular dataset description. Your task is to rate the description on three metrics. \\
    Please make sure you read and understand these instructions carefully. Keep this document open while reviewing, and refer to it as needed. \\
    \\
    \textcolor{teal}{\textbf{Evaluation Criteria:}} \\
    \textbf{1. Completeness (1-10):} \\
    - Evaluates how thoroughly the dataset description covers essential aspects such as the scope of data, query workloads, summary statistics, and possible tasks or applications. \\
    - A high score indicates that the description provides a comprehensive overview, including details on dataset size, structure, fields, and potential use cases. \\
    \\
    \textbf{2. Conciseness (1-10):} \\
    - Measures the efficiency of the dataset description in conveying necessary information without redundancy. \\
    - A high score indicates that the description is succinct, avoiding unnecessary details while employing semantic types (e.g., categories, entities) to streamline communication. \\
    \\
    \textbf{3. Readability (1-10):} \\
    - Evaluates the logical flow and readability of the dataset description. \\
    - A high score suggests that the description progresses logically from one section to the next, creating a coherent and integrated narrative that facilitates understanding of the dataset. \\
    \\
    \textcolor{teal}{\textbf{Evaluation Steps:}} \\
    - Read the dataset description carefully and identify the main topic and key points. \\
    - Assign a score for each criterion on a scale of 1 to 10, where 1 is the lowest and 10 is the highest, based on the Evaluation Criteria. \\
    \\
    \textcolor{teal}{\textbf{Example Evaluations:}} \\
    \textbf{Example 1:} \\
    \textbf{Description:} The dataset provides information on alcohol-impaired driving deaths and occupant deaths across various states in the United States. It includes data for 51 states, detailing the number of alcohol-impaired driving deaths and occupant deaths, with values ranging from 0 to 3723 and 0 to 10406, respectively. Each entry also contains the state abbreviation and its geographical coordinates. The dataset is structured with categorical and numerical data types, focusing on traffic safety and casualty statistics. Key attributes include state names, death counts, and location coordinates, making it a valuable resource for analyzing traffic safety trends and issues related to impaired driving. \\
    \textbf{Evaluation Form (scores ONLY):} \\
    Completeness: 7, Conciseness: 9, Readability: 9 \\
    \\
    \textbf{Example 2:} \\
    \textbf{Description:} The dataset provides a comprehensive overview of traffic safety statistics across various states in the United States, specifically focusing on alcohol-impaired driving deaths and occupant deaths. It includes data from 51 unique states, represented by their two-letter postal abbreviations, such as MA (Massachusetts), SD (South Dakota), AK (Alaska), MS (Mississippi), and ME (Maine). Each entry in the dataset captures critical information regarding the number of alcohol-impaired driving deaths and the total occupant deaths resulting from traffic incidents. \\
    The column "Alcohol-Impaired Driving Deaths" is represented as an integer, indicating the number of fatalities attributed to alcohol impairment while driving. The dataset reveals a range of values, with the highest recorded number being 2367 deaths in Mississippi, highlighting the severity of the issue in certain regions. In contrast, states like Alaska report significantly lower figures, with only 205 alcohol-impaired driving deaths. \\
    The "Occupant Deaths" column also consists of integer values, representing the total number of deaths among vehicle occupants, regardless of the cause. This data spans from 0 to 10406, with Mississippi again showing the highest number of occupant deaths at 6100, which raises concerns about overall traffic safety in the state. \\
    Additionally, the dataset includes a "Location" column that provides geographical coordinates for each state, enhancing the spatial understanding of the data. The coordinates are formatted as latitude and longitude pairs, allowing for potential mapping and geographical analysis of traffic safety trends. \\
    Overall, this dataset serves as a valuable resource for researchers, policymakers, and public safety advocates aiming to understand and address the impact of alcohol on driving safety across different states. It highlights the need for targeted interventions and policies to reduce alcohol-impaired driving incidents and improve occupant safety on the roads. \\
    \textbf{Evaluation Form (scores ONLY):} \\
    Completeness: 8, Conciseness: 7, Readability: 8 \\
    \\
    \textcolor{teal}{\textbf{Final Evaluation Form:}} \\
    Please provide scores for the given dataset description based on the Evaluation Criteria. Do not include any additional information or comments in your response. \\
    \textbf{Evaluation Form (scores ONLY):} \\
    Completeness: \_\_, Conciseness: \_\_, Readability: \_\_ \\
    \bottomrule
    \end{tabular}
    \caption{Guidelines for evaluating dataset descriptions based on completeness, conciseness, and readability.}
    \label{tab:llm_eval_comp_conc_read}
\end{table*}